\documentclass[sigconf]{acmart}

\usepackage{booktabs} % For formal tables
\usepackage{helvet}  %Required
\usepackage{courier}  %Required
\usepackage{url}  %Required
\usepackage{graphicx}  %Required
\usepackage{bm}
\usepackage{amsthm}
\usepackage{array}
\usepackage{algorithm}
\usepackage{algpseudocode}
\usepackage{amsmath}
\usepackage{amssymb}
\usepackage{balance}
\usepackage{bm}
\usepackage{color}
\usepackage{etex}
\usepackage{graphicx}
\usepackage{multirow}
\usepackage{stmaryrd}
\usepackage{subcaption}
\usepackage{url}
\usepackage{thmtools}
\usepackage{mathtools}
\usepackage{todonotes}
\usepackage{pbox}
\usepackage{bbold}
\usepackage{balance}

\newcommand{\squishlist}{
    \begin{list}{$\bullet$}
    { \setlength{\itemsep}{0pt}
        \setlength{\parsep}{3pt}
        \setlength{\topsep}{3pt}
        \setlength{\partopsep}{0pt}
        \setlength{\leftmargin}{1.5em}
        \setlength{\labelwidth}{1em}
        \setlength{\labelsep}{0.5em} } }

\newcommand{\squishlisttwo}{
    \begin{list}{$\bullet$}
        { \setlength{\itemsep}{0pt}
            \setlength{\parsep}{0pt}
            \setlength{\topsep}{0pt}
            \setlength{\partopsep}{0pt}
            \setlength{\leftmargin}{2em}
            \setlength{\labelwidth}{1.5em}
            \setlength{\labelsep}{0.5em} } }

\newcommand{\squishend}{
    \end{list}  }

\setcopyright{rightsretained}
\newcommand{\eg}{\textit{e.g.,}~}

\newcommand{\etc}{\textit{etc.}~}

\newcommand{\mc}{\mathcal}

\newcolumntype{L}{>{$}l<{$}}
\newcolumntype{C}{>{$}c<{$}}
\newcolumntype{R}{>{$}r<{$}}

\copyrightyear{2019}
\acmYear{2019} 
\setcopyright{iw3c2w3}
\acmConference[WWW '19]{Proceedings of the 2019 World Wide Web Conference}{May 13--17, 2019}{San Francisco, CA, USA}
\acmBooktitle{Proceedings of the 2019 World Wide Web Conference (WWW '19), May 13--17, 2019, San Francisco, CA, USA}
\acmPrice{}
\acmDOI{10.1145/3308558.3313460}
\acmISBN{978-1-4503-6674-8/19/05}


\begin{document}
%\fancyhead{}
\title{Signed Distance-based Deep Memory Recommender}

\author{Thanh Tran, Xinyue Liu, Kyumin Lee, Xiangnan Kong}
\affiliation{
	\institution{Department of Computer Science \\ Worcester Polytechnic Institute}
	\state{Massachusetts}
	\country{USA}
}
\email{{tdtran, xliu4, kmlee, xkong}@wpi.edu}	

%\author{Xinyue Liu}
%\affiliation{
%	\institution{Worcester Polytechnic Institute}
%	\state{Massachusetts}
%	\country{USA}
%}
%\email{tdtran@wpi.edu}	
%
%\author{Kyumin Lee}
%\affiliation{
%	\institution{Worcester Polytechnic Institute}
%	\state{Massachusetts}
%	\country{USA}
%}
%\email{tdtran@wpi.edu}
%
%\author{Xiangnan Kong}
%\affiliation{
%	\institution{Worcester Polytechnic Institute}
%	\state{Massachusetts}
%	\country{USA}
%}
%\email{tdtran@wpi.edu}
	
\begin{abstract}
Personalized recommendation algorithms learn a user's preference for an item by measuring a distance/similarity between them. However, some of the existing recommendation models (e.g., matrix factorization) assume a linear relationship between the user and item. This approach limits the capacity of recommender systems, since the interactions between users and items in real-world applications are much more complex than the linear relationship. To overcome this limitation, in this paper, we design and propose a deep learning framework called \emph{Signed Distance-based Deep Memory Recommender}, which captures non-linear relationships between users and items \emph{explicitly} and \emph{implicitly}, and work well in both general recommendation task and shopping basket-based recommendation task. Through an extensive empirical study on six real-world datasets in the two recommendation tasks, our proposed approach achieved significant improvement over ten state-of-the-art recommendation models. %Our source code is available at an anonymized URL.

\end{abstract}

\maketitle
	
\section{Introduction}
\label{sec:introduction}

\begin{figure}[t]
	\centering
	\includegraphics[width=0.325\textwidth]{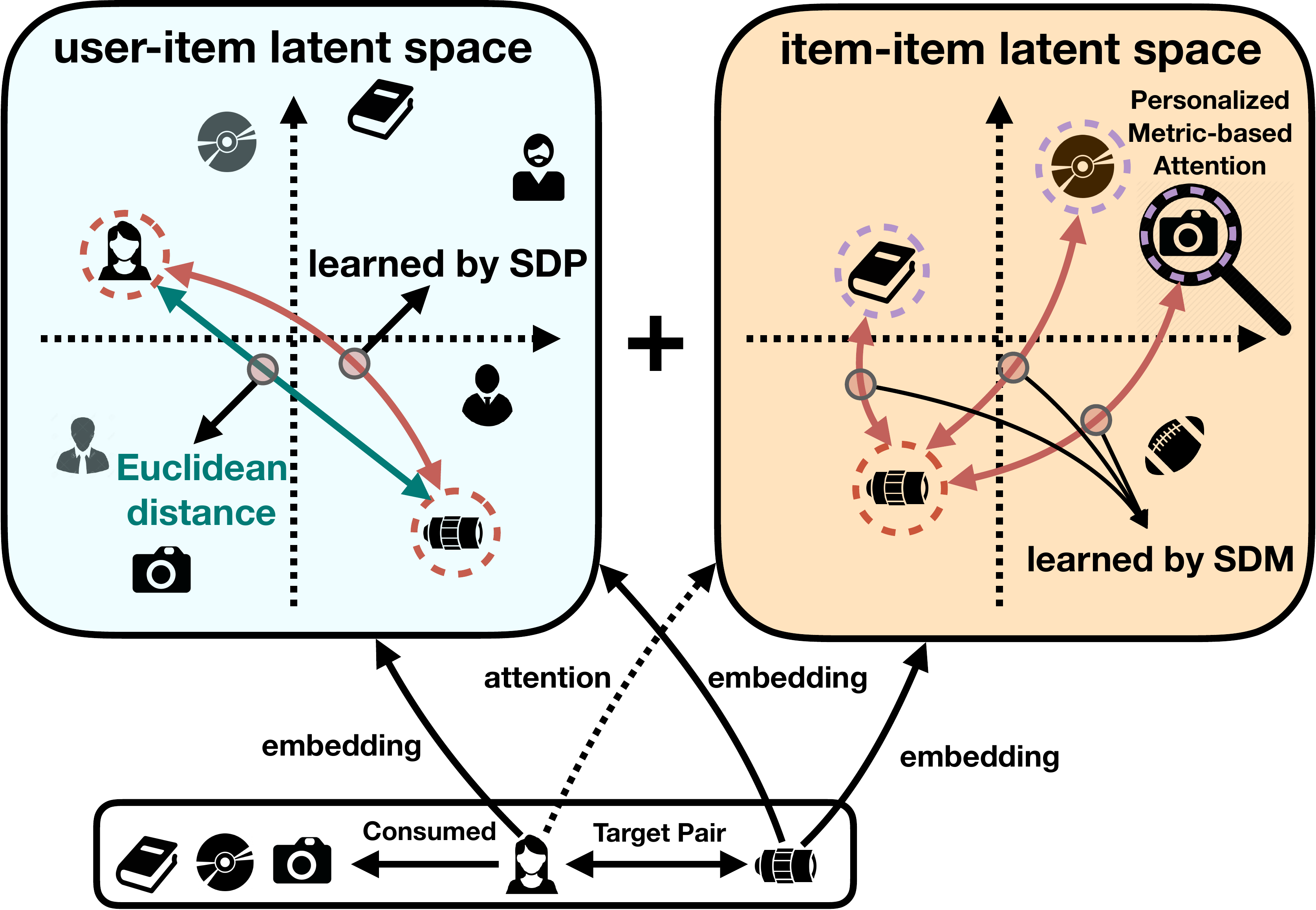}
	\vspace{-5pt}
	\caption{
		We consider a recommender as a signed distance approximator, and decompose the signed distance between a user and an item into two parts:
		the left box learns an explicitly signed distance between the user and item (i.e., the \emph{camera lens}),
		the right box learns an implicitly signed distance between the user and the item via the user's recently consumed items (i.e., the \emph{book}, \emph{CD} and \emph{camera}).
		Our novel personalized metric-based soft attention is applied to the consumed items to optimize their contributions to the output signed distance score.
		Then the two parts are combined to obtain a final score.
		Most of linear latent factor models are equivalent to simply measuring the linear Euclidean distance in the user-item latent space (shown as the green line).
	}
	\label{fig:story}
	\vspace{-15pt}
\end{figure}

%What is the problem?
%Why is it interesting and important?
%Why is it hard? (E.g., why do naive approaches fail?)
%Why hasn't it been solved before? (Or, what's wrong with previous proposed solutions? How does mine differ?)
%What are the key components of my approach and results? Also include any specific limitations.}

Recommender systems \cite{aggarwal2016recommender} have been deployed in many online applications such as e-commerce, music/video streaming services, social media, \etc They have played a vital role for users to explore new items and for companies to increase their revenues. Most of recommendation algorithms model user preferences and item properties based on observed interactions (\eg clicks, reviews, ratings) between users and items \cite{koren2009collaborative,koren2010collaborative,liu2016kernelized}. In a perspective, we can view most of the recommendation models as a measurement of similarity or distance between a user and an item. For instance, the well known latent factor (i.e., matrix factorization) models \cite{koren2008factorization} usually employ an inner product function to approximate the similarity between the user and the item. %By co-registering the embeddings of users and items into a shared latent space, latent factor models assume that if the inner product of the user embedding and the item embedding is large, then the similarity between the corresponding user and item is high. %The high similarity indicates a high likelihood that the user will like the item.
Although the latent factor models achieved competitive performance in some datasets, they did not correctly capture complex (i.e., non-linear) relationships between users and items because the inner product function follows limited linear nature.

Existing recommendation algorithms faced difficulties in finding good kernels for different data patterns \cite{liu2016kernelized}, only focused on user-item latent space without considering the item-item latent space together \cite{he2017neural,liang2018variational,wu2016collaborative,li2015deep,sedhain2015autorec}, or required additional auxiliary information (e.g., item description, music content, reviews) \cite{kim2016convolutional,van2013deep,liu2017deepstyle,chen2017attentive,lu2018coevolutionary}. To overcome the drawbacks, in this paper we aim to propose and build a deep learning framework to learn a non-linear relationship between a user and a target item by measuring a distance from the observed data. In particular, we propose \emph{Signed Distance-based Deep Memory Recommender} (SDMR), which captures non-linear relationship of the user and item \emph{explicitly} and \emph{implicitly}, combines \emph{explicitly} and \emph{implicitly} measured relationship to produce a final distance score for the recommendation, and performs well in both general recommendation task and shopping basket-based recommendation task.

SDMR internally combines two signed distances, each of which is measured by our proposed \emph{Signed Distance-based Perceptron} (SDP) and \emph{Signed Distance-based Memory Network} (SDM). On one hand, SDP explicitly measures a non-linear signed distance between the user and the item. Many existing models \cite{he2016fast,hu2008collaborative} rely on a pre-defined metric such as Euclidean distance (the green line in Figure~\ref{fig:story}) which is much more limited than the customized non-linear signed distance learned from the data (the red curves in Figure~\ref{fig:story}). On the other hand, SDM implicitly measures a non-linear signed distance between the user and the item via the user's recently consumed items. SDM is similar to the item neighborhood-based recommender \cite{sarwar2001item,ning2011slim} in nature. However, it is more advanced in several aspects, as shown in the right side of Figure~\ref{fig:story}. First, SDM only focuses on a set of recently consumed items of the target user (\eg the \emph{book}, \emph{CD} and \emph{camera} in Figure~\ref{fig:story}) as context items. Second, it employs additional memories to learn a novel personalized metric-based attention on the consumed items. The goal of our proposed attention is to compute weights of each consumed item \textit{w.r.t.} the target item (i.e., the \emph{camera lens}). In the example, the attention module assigns higher weights on the \emph{camera} and lower weights on the \emph{book} and \emph{CD}. Unlike our approach, most of the existing neighborhood-based models consider contribution of consumed items to the target item equally, leading to suboptimal results. Last but not the least, we update the attention weights via a gated multi-hop to build a long-term memory within SDM. This multi-hop design helps refine our attention module and produces more accurate attentive scores.

The contributions of this work are summarized as follows:
%\begin{itemize}
\squishlist
	\item We design a deep learning framework which can tackle both general recommendation task and shopping basket-based recommendation task.
	%\item We formalize the recommendation problem as the compound of two separate signed distance measurements.
    \item We propose SDMR that combines two signed distance scores internally measured by SDP and SDM, which capture non-linear relationship between a user and an item explicitly and implicitly.
	\item To better balance the weights among consumed items of the user, we propose a novel multi-hop memory network with a personalized metric-based attention mechanism in SDM.
	\item Extensive experiments on six datasets in two different recommendation tasks demonstrate the effectiveness of our proposed methods against ten baselines.
\squishend

\section{Related Work}
%Recommendation methods can be roughly classified into several categories, such as neighborhood-based models, latent factor models, deep learning models \etc
%We briefly discuss the relation and difference between models in each category and our proposed method respectively in this section.
%\subsubsection{Latent Factor Models}
%\smallskip		
%\noindent\textbf{Latent Factor Models.}
Latent Factor Models (LFM) have been extensively studied in the literature, %\cite{srebro2003weighted,srebro2005maximum,rennie2005fast,mnih2008probabilistic}.
which include Matrix Factorization \cite{hu2008collaborative}, Bayesian Personalized Ranking \cite{rendle2009bpr}, fast matrix factorization for implicit feedbacks (eALS) \cite{he2016fast}, \etc
%These models work by factorizing user-item interactions into user latent space and item latent space using the inner dot product.
Despite their success, LFM suffer from several limitations.
First, LFM overlook associations between the user's previously consumed items and the target item (e.g. mobile phones and phone cases).
Second, LFM usually rely on inner product function, whose linearity limits the capability of modeling complex user-item interactions.
To address the second issue, several non-linear latent factor models have been proposed, with the help of Gaussian process \cite{lawrence2009non} or kernels \cite{liu2016kernelized,zhou2012kernelized}.
However, they either require expensive hyper-parameter tuning or face difficulties in finding good kernels for different data patterns.
%For example, utilized Gaussian process to introduce non-linear capabilities. \cite{} introduced different non-linear kernels for matrix factorization.
%But expensive hyper-parameter tuning processes, and difficulties in making \emph{deeper} models for handling even more complex user-item interactions.
%Besides, they usually require rich content or side information to alleviate the sparsity issue for kernel.
%Our proposed models overcome these issues by leveraging the necessaries of user's consumed items in predicting her next item and exploiting neural net architectures, which are notable for easily adding more layers and integrating non-linear activation functions.

%\subsubsection{Neighborhood-based Models}
Neighborhood-based models \cite{sarwar2001item,ning2011slim} are usually based on the principle that similar users prefer similar items. %and similar items are preferred by similar users.
The problem turns into finding the neighbors of a user or an item based on a pre-defined distance/similarity metric, such as \emph{cosine vector similarity} \cite{lang1995newsweeder,billsus2000user}, \emph{Person Correlation similarity} \cite{deshpande2004item}, \etc
The recommendation quality highly depends on a chosen metric, but finding a good pre-defined metric is usually very challenging.
Furthermore, these models are also sensitive to the selection of neighbors.
%Some hybrid models that combined neighborhood-based models and LFM were proposed to enhance predictive capabilities such as SVD++ \cite{koren2010collaborative}, Factorization Machine\cite{rendle2010factorization}.
%However, these methods still suffer from linear computations using the inner dot product. Additionally, while the different user's consumed items can contribute differently to the predicting item, those methods assign the contributions equally, leading to suboptimal results.
Our proposed SDM is similar to neighborhood-based models in nature, but it exploits a novel personalized metric-based attention for assigning attentive weights to context items. Therefore, our approach is more robust and less sensitive than conventional neighborhood-based models.

%As one of the most widely used collaborative filtering technique,  latent factor models or matrix factorization has been extensively studied in the literature \cite{srebro2003weighted,srebro2005maximum,rennie2005fast,mnih2008probabilistic}.
%These models work by co-registering the user embeddings and item embeddings into a shared latent space, and employing inner product function to approximate the interactions between users and items.
%Since the observed interactions are usually sparse, latent factor models can be optimized more efficiently than neighborhood-based models via gradient descent, thus attracted extensive attention in the last decade.
%As discussed before, most of latent factor models suffer from the limitation of linearity lies in the nature of inner product function.
%Toward this end, several non-linear latent factor models have been proposed.
%\cite{lawrence2009non} proposed to utilize Gaussian process to introduce non-%linearity, but it is highly expensive to perform the hyper-parameter tuning for it.
%\cite{liu2016kernelized,zhou2012kernelized} proposed to make the factorization non-linear by incorporating kernels.
%However, it is difficult and time consuming to find good kernels for the data.
%Besides, they usually require rich content or side information to alleviate the sparsity issue for kernel.
%Intuitively, our proposed method can be viewed as a more generalized version of latent factor models as shown in Fig.~\ref{fig:story}.

%\subsubsection{Deep Learning Models}
NeuMF \cite{he2017neural} is a neural network that generalizes matrix factorization via Multi Layer Perceptron (MLP) for learning non-linear interaction functions.
Similarly, some other works \cite{liang2018variational,wu2016collaborative,li2015deep,sedhain2015autorec} substitute MLP with auto-encoder architecture.
It is worth noting that all these approaches are limited by only considering the user-item latent space, and overlook the correlations in the item-item latent space.
%Auto-encoder is another widely adopted approach \cite{liang2018variational,wu2016collaborative,li2015deep,sedhain2015autorec}, which .
%It works by decomposing the user-item interaction matrix with an encoder followed by a decoder for reconstruction.
%Even though these model architectures adopted non-linear activation functions to overcome the linear issues of traditional recommendation systems, they still ignored the connection between the predicting item and the context items.
%In contrast, we consider to model not only the explicit user's preference on the predicting item, but also the effect of the context items toward the predicting items in a non-linear fashion.
% Brief review since we do not consider auxiliary data.
Besides, some deep learning based works \cite{lu2018convolutional,tay2018multi,seo2017interpretable,ma2019gated,ma2018point} employ auxiliary information such as item description \cite{kim2016convolutional}, music content \cite{van2013deep}, item visual features \cite{liu2017deepstyle,chen2017attentive}, reviews \cite{lu2018coevolutionary} to address the cold-start problem.
However, this auxiliary information is not always available, and it limits their applicability in many real-world systems.
Another line of works use deep neural networks to model temporal effects of consumed items \cite{hidasi2015session,wu2017recurrent,quadrana2017personalizing,tang2018personalized}.
Although our proposed methods do not explicitly consider the temporal effects, SDM utilizes the time information to select a set of recently consumed items as the context items of the target item.
%which enables the ``local temporal awareness" for users.

%\cite{} proposed to use a Hierarchical Recurrent Neural network for personalized session-based recommendations. \cite{} considered users' consumed items as user sequences and utilized CNN architecture for the personalized sequential recommendation. Different from those works, we do not model the order of items.

%In this category, there existed a majority of works exploited CNN based architecture, RNN based architecture \cite{}, autoencoder based \cite{wang2015collaborative}.
%However, we only used an user-item-preference matrix without requiring auxiliary information. Incorporating additional side information can further improve the model's performance, but it is not the scope of our work in this paper.

%First, both SDM and CMN++ exploited memory network for collaborative filtering recommendations. However, we note several key differences between our design and CMN as: (i) our proposed SDM followed item-based neighborhood design, which was shown to perform slightly better than user-based neighborhood design \cite{linden2003amazon,sarwar2001item}; (ii) SDM used our proposed first-attempt personalized metric-based attention mechanism and produces signed distance scores as output, whereas CMN barely exploited a traditional inner product based attention; and (iii) our multi-hop design exploited gated multi-hop design, which was show to perform better than original multi-hop design \cite{liu2017gated}.

The most closely related work to our work is recently proposed (\emph{Collaborative Memory Network} (CMN) \cite{ebesu2018collaborative}).
In this work, Memory Network \cite{sukhbaatar2015end} is adapted to measure similarities between users and user neighbors. Key differences between our work and CMN are as follows: (i) First, we follow an item neighborhood based design, whereas CMN follows a user neighborhood based design. The prior work showed that item neighborhood based models slightly outperformed user neighbor based models \cite{linden2003amazon,sarwar2001item}; (ii) Second, our proposed SDM model uses our proposed personalized metric-based attention mechanism and produces signed distance scores as output, whereas CMN exploited a traditional inner product based attention; (iii) Third, we use a gated multi-hop architecture \cite{liu2017gated}, which was shown to perform better than the original multi-hop design \cite{sukhbaatar2015end}. % showed that a gated multi-hop design outperformed the original multi-hop design.

%However, they still used the inner product for assessing weights of item pairs in the attention module.
%We fully exploit deep metric learning in all our components to further extend the capability.
%Moreover, we adopted a Gated Memory Network \cite{liu2017gated}, which performed better than original Memory Network \cite{sukhbaatar2015end}.

%\cite{he2017neural} proposed a neural network architecture to generalize the latent factor models.

%VAE recommender \cite{liang2018variational}:
%Memory network \cite{ebesu2018collaborative}:
%Attention models: \cite{tay2018multi,tay2018latent}

%\subsubsection{Temporal Recommender}
%Temporal recommender \cite{koren2009collaborative,xiong2010temporal,liu2017bicycle} incorporates the temporal effects by considering the order of interactions.
%%TimeSVD++ \cite{} is a latent factor model that uses linear functions to approximate the temporal dynamics of users and items.
%%\cite{} proposed a tensor factorization model for the temporal recommender.
%%with Bayesian treatment to avoid hyper-parameter tunning.
%%\cite{} attempted to advance the performance with the concept of life cycles instead of wall-time dynamics.
%Recently, RNN-based solutions were also explored \cite{wu2017recurrent} for this direction.
%Although our proposal does not explicitly consider the temporal effects, but SDM utilizes the time information to select %the most recently consumed items, which enables the ``local temporal awareness" for users.

% Reserved for the extended version
%\subsubsection{Memory Network}

\section{Problem Statement}
In this section, we describe two recommendation problems: (i) general recommendation task; and (ii) shopping basket-based recommendation task. In following sections, we focus on solving them.
%Before going into details of our proposed models, we first define two problems that our proposed models will solve as follows:

\noindent\textbf{General recommendation task:} Given a whole item set $V = \{ v_1, v_2, ..., v_{|V|}\}$, and a whole user set $U = \{u_ 1, u_2, ..., u_{|U|}\}$. Each user $u_i \in U$ may consume several items $\{v_{i1}, v_{i2}, ..., v_{ik}\}$ in $V$, denoted as a set of context items $c$. In this task, given previously consumed items of a user $u_i$, a recommendation model predicts a next target item $v_j$ that $u_i$ may prefer, denoting this task as estimating $P(u_i, v_j|c)$. Note that some existing works assume independent relationships between $v_j$ and context items in the set $\boldsymbol{c}$, leading to $P(u_i, v_j|c) = P(u_i, v_j)$ \cite{he2017neural,he2016fast}. In our work, we model the $u_i$'s preference on $v_j$ in two steps: (i) an explicit preference of $u_i$ on $v_j$ in a signed distance based perceptron, and (ii) an implicit preference of $u_i$ on $v_j$ via summing attentive effects of context items toward target item $v_j$ in a signed distance based memory network.

\noindent\textbf{Shopping Basket-based recommendation task:} This problem is based on the fact that users go shopping offline/online and add some items into a basket/cart together. Each shopping basket/cart is seen as a transaction, and each user may shop once or multiple times, leading to one or multiple transactions. Let $T^{(u)} = \{t_1, t_2, ..., t_{|T^{(u)}|}\}$ as a set of the user $u$'s transactions, where $|T^{(u)}|$ denotes the number of user $u$'s transactions. Each transaction $t_i = \{v_1, v_2, ..., v_{|t_i|}\}$ consists of several items in the whole item set $V$. In this problem, it is assumed that all the items in $t_i$ are inserted into the same basket at the same time, ignoring the actual order of the items being inserted and considering $t_i$'s transaction time as each item's insertion time. Given a target item $v_j \in t_i$, the rest of the items in $t_i$ will be seen as the context items of $v_j$, denoted as $c$ (i.e. $c=t_i\textbackslash \{v_j\}$). Then, given the set of context items $c$, a recommendation model predicts a conditional probability $P(u, v_j | c)$, which is interpreted as the conditional probability that $u$ will add the item $v_j$ into the same basket with the other items $c$.

%Both two recommendation tasks above are popular in the literature \cite{quadrana2017personalizing,rendle2010factorizing,feng2015personalized,he2017neural}. The \emph{general recommendation task} differs from the \emph{shopping basket-based recommendation task} because there is no specific context items of the target item in the \emph{general recommendation task}. %However, if we merge all consumed items of each user into one transaction, then the \emph{general recommendation task} can be seen as a \emph{shopping basket-based recommendation task} with only one long transaction per each user.
%Without special mention, the two tasks are \emph{personalized} recommendation problems. In fact, there existed \emph{non-personalized} recommendation problems such as \emph{session-based recommendation} \cite{hidasi2015session} without assigning any user to any transaction (an example dataset in this type is \emph{YooChoose} \footnote{https://recsys.acm.org/recsys15/challenge/}). However, we are interested in personalized problems because they are more preferred \cite{quadrana2017personalizing,rendle2010factorizing,feng2015personalized}.

Both of the recommendation tasks above are popular in the literature \cite{quadrana2017personalizing,rendle2010factorizing,feng2015personalized,he2017neural}. The \emph{general recommendation task} differs from the \emph{shopping basket-based recommendation task} because there is no specific context items of the target item in the \emph{general recommendation task}. %However, if we merge all consumed items of each user into one transaction, then the \emph{general recommendation task} can be seen as a \emph{shopping basket-based recommendation task} with only one long transaction per each user.
Note that the two tasks are \emph{personalized} recommendation problems. In fact, there are \emph{non-personalized} recommendation problems such as \emph{session-based recommendation} \cite{hidasi2015session}, where users (i.e. user IDs) are not available in transactions. %\footnote{(an example dataset in this type is \emph{YooChoose}} \footnote{https://recsys.acm.org/recsys15/challenge/}).
%However, we are interested in personalized problems because they are more preferred \cite{quadrana2017personalizing,rendle2010factorizing,feng2015personalized}.
However, in this paper, we focus on \emph{personalized} recommendation tasks because they are more preferred %than non-personalized recommendation tasks
in the literature \cite{quadrana2017personalizing,rendle2010factorizing,feng2015personalized}.

%The \emph{general recommendation task} is more general than the \emph{shopping basket-based recommendation task} because there is no specific context items of the target item in the \emph{general recommendation task}. However, if we merge all of the consumed items of a user into one transaction, then the \emph{general recommendation task} can be seen as a \emph{shopping basket-based recommendation task} with only one transaction per user. In this paper, we focus on \emph{personalized} recommendation tasks \cite{quadrana2017personalizing,rendle2010factorizing,feng2015personalized}. %Note that there are also \emph{non-personalized} recommendation tasks such as the \emph{session-based recommendation} problem \cite{hidasi2015session} without assigning any user to any transaction (an example dataset in this type is \emph{YooChoose} \footnote{https://recsys.acm.org/recsys15/challenge/}). %However, we are interested in personalized problems because they are more preferred .

\section{Proposed Methods}
Our proposed \emph{Signed Distance-based Deep Memory Recommender} (SDMR) consists of two major components: \emph{Signed Distance-based Perceptron} (SDP) and \emph{Signed Distance-based Memory network} (SDM). We first describe an overview of our models as follows:
\squishlist
\item Given a target user $i$ and a target item $j$ as two one-hot vectors, we pass the two vectors through the user and item embedding spaces to get user embedding $u_i$ and item embedding $v_j$.
\item On one hand, our proposed \emph{Signed Distance-based Perceptron} (SDP) will measure a signed distance score between $u_i$ and $v_j$ by a multi-layer perceptron network.
\item On the other hand, given target user $i$, target item $j$, and the user $i$'s recently consumed context items $s$ as the input, our \emph{Signed Distance-based Memory network} (SDM) will measure a signed distance score between user $i$ and item $j$ via attentive distances between context items $s$ and target item $j$.
\item Then, the \emph{Signed Distance-based Deep Memory Recommender} (SDMR) model will measure a total distance between user $i$ and item $j$ by learning a combination of SDP and SDM. The smaller the total distance is, the more likely user $i$ will consume item $j$.
\squishend

Next, we describe SDP, SDM, and SDMR in detail.	

\vspace{-5pt}	
\subsection{Signed Distance-based Perceptron (SDP)}
\begin{figure}
	\centering
	\includegraphics[width=0.25\textwidth]{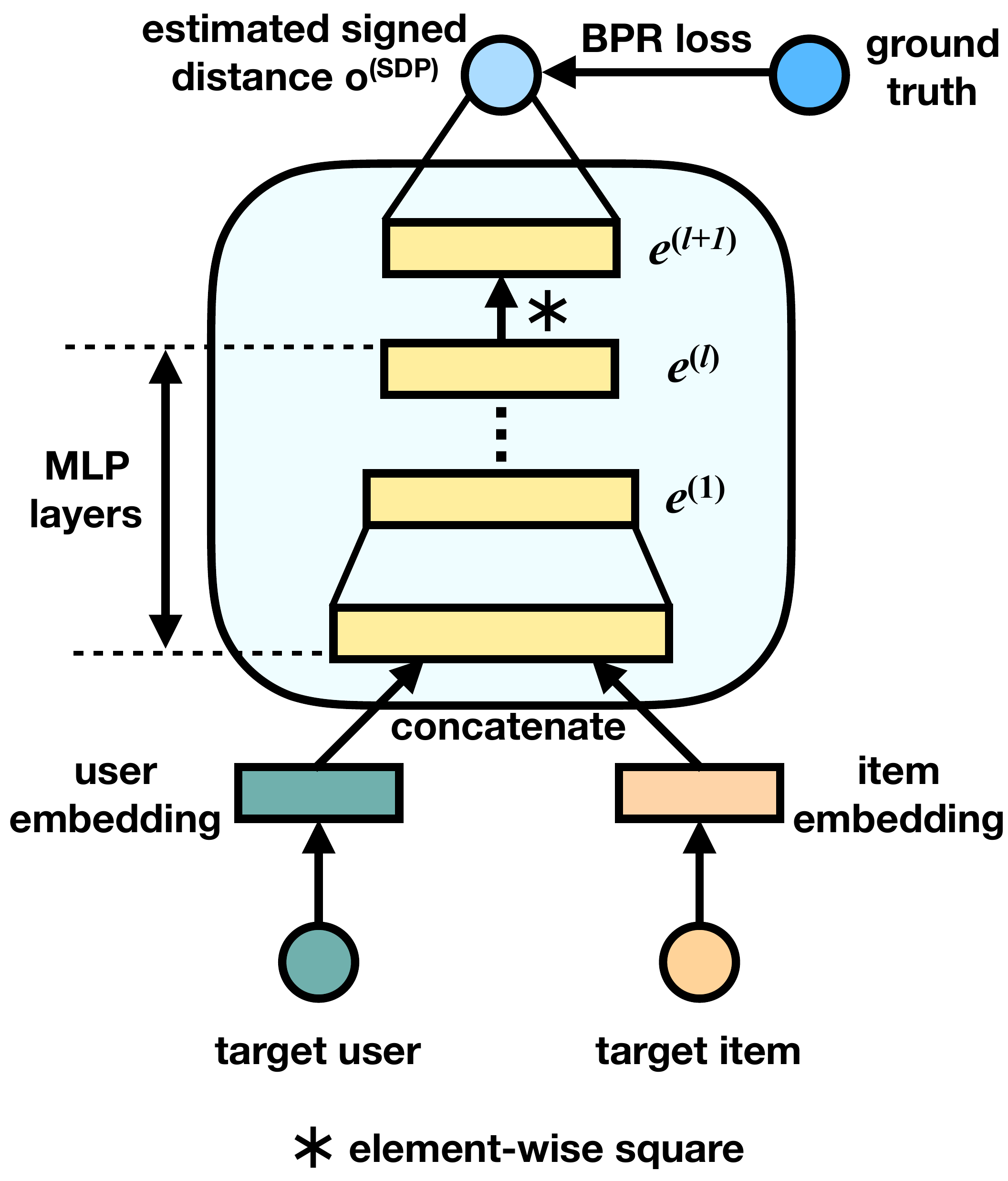}
	\vspace{-10pt}
	\caption{
		The illustration of our SDP model.
		%The user and item embeddings are first concatenated. Next, the concatenated embeddings are modeled through a MLP. Then, an element-wise square function is applied on the output vector of the MLP. Finally, a fully connected layer will combine different dimensions in the output vector.
	}
	\vspace{-10pt}
	\label{fig:GMP}
\end{figure}

We first propose \emph{Signed Distance-based Perceptron} (SDP) that \emph{explicitly} learns a signed distance between a target user $i$ and a target item $j$. An illustration of SDP is shown in Figure~\ref{fig:GMP}.
%Specifically, SDP is designed to learn a signed distance between a given target user $i$ and a given target item $j$.
%and it reduces to conventional matrix factorization under specific configuration.
Let the embedding of a target user $i$ be $\boldsymbol{u}_i \in \mathbb{R}^{d}$, and the embedding of a target item $j$ be $\boldsymbol{v}_j \in \mathbb{R}^{d}$, where $d$ is the number of dimensions in each embedding.
First, SDP takes a concatenation of these two embeddings as the input and proceeds as follows:
\vspace{-5pt}
\begin{align}
\label{eq:gmp-layer1} &\boldsymbol{e}^{(1)} = f_1(\mathbf{W}^{(1)} \begin{bmatrix} \boldsymbol{u}_i \\ \boldsymbol{v}_j \end{bmatrix} + \boldsymbol{b}^{(1)}) \\
\label{eq:gmp-layer2} &\boldsymbol{e}^{(2)} = f_2(\mathbf{W}^{(2)} \boldsymbol{e}^{(1)} + \boldsymbol{b}^{(2)}) \\
\label{eq:gmp-layer3} &\cdots \\
\label{eq:gmp-layer4} &\boldsymbol{e}^{(\ell)} = f_\ell(\mathbf{W}^{(\ell)} \boldsymbol{e}^{(\ell-1)} + \boldsymbol{b}^{(\ell)}) \\
%\label{eq:gmp-layerl} &\boldsymbol{e}^{(\ell+1)} =  f_{\ell+1}({\boldsymbol{e^{(\ell)}}}^{2}) \\
\label{eq:gmp-layerl} &\boldsymbol{e}^{(\ell+1)} =  square({\boldsymbol{e^{(\ell)}}}) \\
\label{eq:gmp-layero} &o^{(SDP)} = {\boldsymbol{w}^{(o)}}^{\top} \boldsymbol{e}^{(\ell+1)} + \boldsymbol{b}^{(o)}
\end{align}
where $f_l(\cdot)$ refers to a non-linear activation function at the layer $l^{th}$ (e.g. \texttt{sigmoid}, \texttt{ReLu} or \texttt{tanh}), and $square(\cdot)$ denotes an element-wise square function (e.g $square([2, 3]) = [6, 9]$). Through experimental results, we choose \texttt{tanh} as the activation function because it yields slightly better results than \texttt{ReLu}. From now on, we will use $f(\cdot)$ to denote the \texttt{tanh} function.
It can be easily observed that Eq.~(\ref{eq:gmp-layer1}) -- (\ref{eq:gmp-layer4}) form a trivial Multi-Layer Perceptron (MLP) network, which is a popular design \cite{he2017neural,xue2017deep} to learn a complex and non-linear interaction between user embedding $\boldsymbol{u}_i$ and item embedding $\boldsymbol{v}_j$. Our new design starts at Eq.~(\ref{eq:gmp-layerl}) -- Eq.~(\ref{eq:gmp-layero}). % After getting a non-linear interaction vector $\boldsymbol{e}^l$ between $\boldsymbol{u}_i$ and $\boldsymbol{v}_j$ via a MLP
In Eq.~(\ref{eq:gmp-layerl}), we apply the element-wise squared function $square(\cdot)$ to the output vector $\boldsymbol{e}^{(l)}$ of the MLP and obtain a new output vector $\boldsymbol{e}^{(l+1)}$. Next, in Eq.~(\ref{eq:gmp-layero}), we use a fully connected layer $\boldsymbol{w}^{(o)}$ to combine different dimensions in $\boldsymbol{e}^{(l+1)}$ and yields a final distance value $o^{(SDP)}$. Our idea of using $\boldsymbol{w}^{(o)}$ in here is that after applying the element-wise square function $square(\cdot)$ in Eq.~(\ref{eq:gmp-layerl}), all the dimensions in $\boldsymbol{e}^{(l+1)}$ will be non-negative. Thus, we consider each dimension of $\boldsymbol{e}^{(l+1)}$ as a distance value. The edge weights $\boldsymbol{w}^{(o)}$ will then be used to combine those distant dimensions to provide a more fine-grained distance.
%The non-layer function $g(\cdot)$ performs the \emph{power-of-n} for all input elements\footnote{Increase/decrease n will increase/decrease the distance margin. In general, $g(\cdot)$ generalizes Minkowski distance.}.
%We set \emph{n=2} to produce a squared Euclidean distance.

%($o \ge 0$).
%Eq.~(\ref{eq:gmp-layerl}) takes the element-wise square of $\ell$-layer embedding, and Eq.~(\ref{eq:gmp-layero}) is the output layer that yields the final distance score.
%The non-layer activation function $f(\cdot)$ could be chosen accordingly based on the characteristics of dataset.

We note that SDP can be reduced to a squared Euclidean distance with the following setting: at Eq.~(\ref{eq:gmp-layer1}), $\mathbf{W}^{(1)} = [\mathbb{1}, -\mathbb{1}]$ with $\mathbb{1}$ denotes an identity matrix and so $\mathbf{W}^{(1)} \begin{bmatrix} \boldsymbol{u}_i \\ \boldsymbol{v}_j \end{bmatrix} = \boldsymbol{u}_i - \boldsymbol{v}_j$; the activation $f(\cdot)$ is an identity function; the number of MLP layers $\ell = 1$; the edge-weights layer at Eq.~(\ref{eq:gmp-layero}): $\boldsymbol{w}^{(o)} = \boldsymbol{1}$ (e.g. the all-ones matrix), bias $\boldsymbol{b}^{(o)} = \boldsymbol{0}$.
Note that if $\boldsymbol{w}^{(o)}$ in Eq.~(\ref{eq:gmp-layero}) is an all-negative layer, it will yield a negative value, which we name as a signed distance\footnote{https://en.wikipedia.org/wiki/Signed\_distance\_function} score. If we see each user $\boldsymbol{i}$ as a point in multi dimensional space, and the user's preference space is defined by a boundary $\Omega$, we can interpret this signed distance score as follows: When the item $\boldsymbol{j}$ is out of the user $\boldsymbol{i}$'s preference boundary $\Omega$, the distance $d(\boldsymbol{i}, \boldsymbol{j})$ between them is positive (i.e. $d(\boldsymbol{i}, \boldsymbol{j})$ > 0) and it reflects that user $\boldsymbol{i}$ does not prefer item $\boldsymbol{j}$. When the distance between user $\boldsymbol{i}$ and item $\boldsymbol{j}$ is shortened and $\boldsymbol{j}$ is right on the boundary $\Omega$, the distance between them is zero and it indicates user $\boldsymbol{i}$ likes item $\boldsymbol{j}$. As $\boldsymbol{j}$ is coming inside $\Omega$, the distance between them becomes negative and reflects a higher preference of user $\boldsymbol{i}$ on item $\boldsymbol{j}$. In short, we can see SDP as a signed distance function, which could learn a complex signed distance between a user and an item via a MLP architecture with non-linear activations and an element-wise square function $square(\cdot)$.
%	\todo[inline]{applications of signed distance? cite in here}
In the recommendation domain, the signed distances will provide more fine-grained distance values, thus, reflecting users' preferences on items more accurately.% (i.e. accurately rank items for the user).

\subsection{Signed Distance-based Memory Network (SDM)}
\begin{figure}
	\centering
	\includegraphics[width=0.44\textwidth]{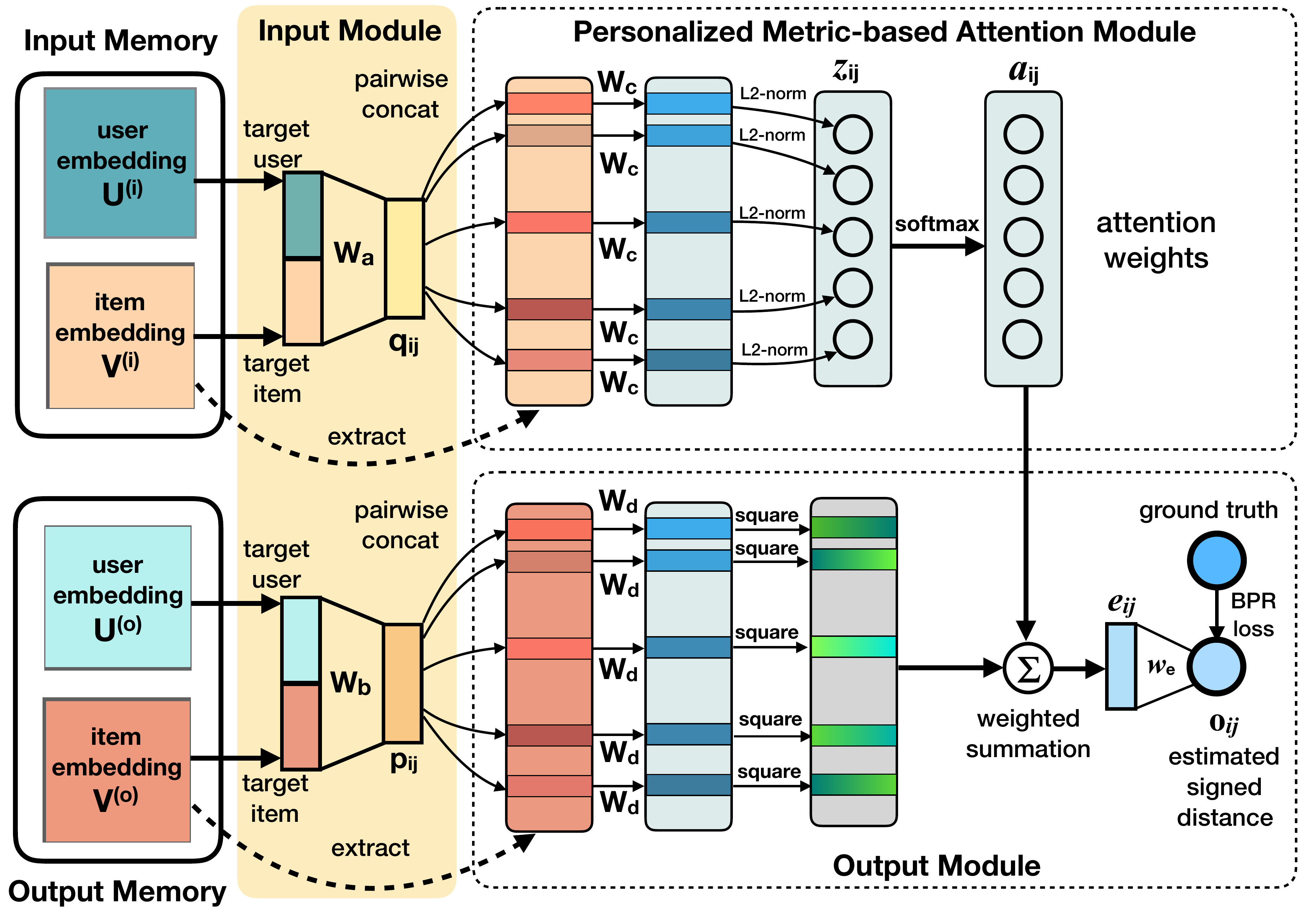}
	%\vspace{-15pt}
	\vspace{-10pt}
	\caption{
		The illustration of a single-hop SDM, which consists of a memory module, an input module, an attention module, and an output module.
		%Its \emph{memory module} has separated input, output memory. The \emph{input module} will form a searching query. The \emph{personalized distance-based attention module} will assign (\emph{high}) attentive scores to (\emph{matching}) context items. The \emph{output module} produces a final attentive signed distance score.
	}
	\label{fig:MMN}
	\vspace{-10pt}
\end{figure}
We propose a multi-hop memory network, \emph{Signed Distance-based Memory network} (SDM), to model \emph{implicit} preference of a user on the target item via the user's previously consumed items (i.e., context items). The implicit preference is represented as a signed distance. First, we describe a single-hop SDM, and then describe how to extend it into a multi-hop design. Following the traditional architecture of a memory network \cite{sukhbaatar2015end,liu2017gated,xiong2016dynamic}, our proposed single-hop SDM has four main components: a memory module, an input module, an attention module, and an output module. The overview of SDM's architecture is presented in Figure~\ref{fig:MMN}. We will go into details of each SDM's module as follows:

\subsubsection{Memory Module:}
We maintain two memories called input memory and output memory.
The input memory contains two embedding matrices $\mathbf{U}^{(i)} \in \mathbb{R}^{M \times d}$ and $\mathbf{V}^{(i)} \in \mathbb{R}^{N \times d}$, where $M$ and $N$ are the number of users and the number of items in the system, respectively. $d$ denotes the embedding size of each user and each item.
Similarly, the output memory also contains two embedding matrices $\mathbf{U}^{(o)} \in \mathbb{R}^{M \times d}$ and $\mathbf{V}^{(o)} \in \mathbb{R}^{N \times d}$.
As shown in Figure~\ref{fig:MMN}, the input memory will be used to calculate attention weights of a user's consumed items (i.e., context items), whereas the output memory will be used to measure a final signed distance between the target user and the target item via the user's context items.

Given a target user $i$, a target item $j$ and a set of user $i$'s consumed items as context items $\mathcal{T}^{i}_j$, the output of this module is the embeddings of user $i$, item $j$, and all context items $k \in \mathcal{T}^{i}_j$: ($\boldsymbol{u}_i, \boldsymbol{v}_j,$ <$\boldsymbol{v}_1, \boldsymbol{v}_2, ..., \boldsymbol{v}_k$>). Since this module has a separated input memory and output memory, we obtain ($\boldsymbol{u}_i^{(i)}, \boldsymbol{v}_j^{(i)},$ <$\boldsymbol{v}_1^{(i)}, \boldsymbol{v}_2^{(i)}, ..., \boldsymbol{v}_k^{(i)}$>) as the output of the input memory, and ($\boldsymbol{u}_i^{(o)}, \boldsymbol{v}_j^{(o)},$ <$\boldsymbol{v}_1^{(o)}, \boldsymbol{v}_2^{(o)}, ..., \boldsymbol{v}_k^{(o)}$>) as the output of the output memory. It is obvious that $\boldsymbol{u}_i^{(i)}$ is the $i$-th row of $\mathbf{U}^{(i)}$, $\boldsymbol{v}_j^{(i)}$ and $\boldsymbol{v}_k^{(i)}$ are the corresponding $j$-th and $k$-th row of $\mathbf{V}^{(i)}$. A similar explanation is applied to $\boldsymbol{u}_i^{(o)}$ $\boldsymbol{v}_j^{(o)}$, and $\boldsymbol{v}_k^{(o)}$.

\subsubsection{Input Module:}
The goal of the input module is to form a non-linear combination between the target user embedding and the target item embedding. Given the target user embedding $\boldsymbol{u}_i^{(i)}$ and the target item embedding $\boldsymbol{v}_j^{(i)}$ from the input memory in the memory module, following the widely adopted design in multimodal deep learning work \cite{zhang2014start,srivastava2012multimodal}, the input module simply concatenates the two embeddings, and then applies a fully connected layer with a non-linear activation $f(\cdot)$ (i.e. \texttt{tanh} function) to obtain a coherent hidden feature vector as follows:
%\vspace{-5pt}
\begin{align}
\label{eq:MMN_input_module}
\boldsymbol{q}_{ij} = f \Big( \mathbf{W}_a
\begin{bmatrix}
\boldsymbol{u}_i^{(i)}\\
\boldsymbol{v}_j^{(i)}
\end{bmatrix} + \boldsymbol{b}_a
\Big)
\end{align}
where $\mathbf{W}_a \in \mathbb{R}^{d \times 2d}$ is the weights of input module. Note that $q_{ij} \in \mathbb{R}^d$ can be seen as a query embedding in Memory Network \cite{sukhbaatar2015end}.

Similarly, if the inputs of the input module are the target user embeddings $\boldsymbol{u}_i^{(o)}$ and the target item embeddings $\boldsymbol{v}_j^{(o)}$ from the output memory, we can form a non-linear combination between $\boldsymbol{u}_i^{(o)}$ and $\boldsymbol{v}_j^{(o)}$ (i.e. an output query), denoted as $\boldsymbol{p}_{ij}$, as follows:
\vspace{-5pt}
\begin{align}
\label{eq:MMN_input_module_p}
\boldsymbol{p}_{ij} = f \Big( \mathbf{W}_b
\begin{bmatrix}
\boldsymbol{u}_i^{(o)}\\
\boldsymbol{v}_j^{(o)}
\end{bmatrix} + \boldsymbol{b}_b
\Big)
\end{align}
%and we omit the bias term throughout this paper for simplicity.

\subsubsection{Attention Module:}
The goal of the attention module is to assign attentive scores to different context items (or candidates) given the combined vector (or a query) $\boldsymbol{q}_{ij}$ of the target user $i$ and target item $j$ obtained in Eq.~(\ref{eq:MMN_input_module}).
First, we calculate the squared $\mathcal{L}2$ distance between $\boldsymbol{q}_{ij}$ and each candidate item $\boldsymbol{v}_k^{(i)}$ as follows:
%\begin{align}
%\alpha_{ijk} = \text{exp}\Big(-\Big\Vert{f\Big(\mathbf{W}_b
%    \begin{bmatrix}
%    \boldsymbol{q}_{ij} \\
%    \boldsymbol{v}_k^{(i)}
%    \end {bmatrix} + \boldsymbol{b}_b\Big)}\Big\Vert_2\Big)
%\end{align}

\vspace{-10pt}
\begin{align}
\label{eq:pre-attention}
z_{ijk} =  \Big\Vert {f \Big(\mathbf{W}_c
	\begin{bmatrix}
	\boldsymbol{q}_{ij} \\
	\boldsymbol{v}_k^{(i)}
	\end {bmatrix} +
	\boldsymbol{b}_c
	\Big)}
\Big\Vert _2 ^2
\end{align}
where $||\cdot||_2$ refers to the $\mathcal{L}2$ distance (or Euclidean distance), which is widely used in previous works to measure similarity among items \cite{feng2015personalized} or between users and items \cite{hsieh2017collaborative}. To better understand our intuition in Eq. (\ref{eq:pre-attention}), we will break it into smaller parts and explain them.
First, similar to the intuition of Eq.~(\ref{eq:MMN_input_module}), we have $f\Big(\mathbf{W}_c \begin{bmatrix} \boldsymbol{q}_{ij} \\ \boldsymbol{v}_k^{(i)} \end {bmatrix} + \boldsymbol{b}_c\Big)$ component to define a non-linear combination between the input query $\boldsymbol{q}_{ij}$ and each context item embeddings $\boldsymbol{v_k}^{(i)}$. Then, $||\cdot||_2^2$ will measure the squared $\mathcal{L}2$ distance of the combined vector.
It is worth to note that with a following setting: $\boldsymbol{W}_a$ = $[\boldsymbol{0} , \mathbb{1}]$ where $\mathbb{1}$ refers to an identity matrix and $\boldsymbol{0}$ is an all-zeros matrix; $f(\cdot)$ is an identity function;  $\boldsymbol{W}_c$ = $[\mathbb{1} , -\mathbb{1}]$; bias terms $\boldsymbol{b}_a = \boldsymbol{b}_c = 0$.
Then, in Eq.~(\ref{eq:MMN_input_module}), $\boldsymbol{q}_{ij} = f \Big( \mathbf{W}_a
\begin{bmatrix} \boldsymbol{u}_i^{(i)} \\ \boldsymbol{v}_j^{(i)} \end{bmatrix} + \boldsymbol{b}_a \Big) = \boldsymbol{v}_j^{(i)}$;
in Eq.~(\ref{eq:pre-attention}), $f\Big(\mathbf{W}_c \begin{bmatrix} \boldsymbol{q}_{ij} \\ \boldsymbol{v}_k^{(i)} \end {bmatrix} + \boldsymbol{b}_c\Big) = \boldsymbol{v}_j^{(i)} - \boldsymbol{v}_k^{(i)}$, and $z_{ijk} = ||(\boldsymbol{v}_j^{(i)} - \boldsymbol{v}_k^{(i)})||_2 ^2$, which simply generalizes a squared $\mathcal{L}2$ distance between the target item $j$ and the context item $k$.
Additionally, with another setting: $\boldsymbol{W}_a$ = $[\mathbb{1} , -\mathbb{1}]$; $f(\cdot)$ is an identity function;  $\boldsymbol{W}_c$ = $[\mathbb{1} , \mathbb{1}]$; bias terms $\boldsymbol{b}_a = \boldsymbol{b}_c = 0$.
Then, in Eq.~(\ref{eq:MMN_input_module}), $\boldsymbol{q}_{ij} = f \Big( \mathbf{W}_a
\begin{bmatrix} \boldsymbol{u}_i^{(i)} \\ \boldsymbol{v}_j^{(i)} \end{bmatrix} + \boldsymbol{b}_a \Big) = \boldsymbol{u}_i^{(i)} - \boldsymbol{v}_j^{(i)}$, in Eq.~(\ref{eq:pre-attention}), $f\Big(\mathbf{W}_c \begin{bmatrix} \boldsymbol{q}_{ij} \\ \boldsymbol{v}_k^{(i)} \end {bmatrix} + \boldsymbol{b}_c\Big) = \boldsymbol{u}_i^{(i)} - \boldsymbol{v}_j^{(i)} + \boldsymbol{v}_k^{(i)}$, and $z_{ijk} = ||(\boldsymbol{v}_k^{(i)} + \boldsymbol{u}_i^{(i)} - \boldsymbol{v}_j^{(i)})||_2 ^2$, which simply generalizes a squared $\mathcal{L}2$ distance between the target item $j$ and the context item $k$ where the user $i$ plays as a translator \cite{he2017translation}. The two examples above show that our proposed design can learn a more generalized distance between target and context items.

The output squared $\mathcal{L}2$ distance in Eq.~(\ref{eq:pre-attention}) will show how similar the target item $j$ and the context item $k$ are. The lower the distance score is, the more similar two items $j$ and $k$ are. Next, we use the Softmax function to normalize and obtain attentive score between $j$ and $k$ as follows:
\vspace{-5pt}
\begin{align}
\label{eq:attention}
a_{ijk} = \frac{exp(-z_{ijk})}{\sum_{p \in \mathcal{T}^{i}_j} exp(-z_{ijp})}
\end{align}
where $\mathcal{T}^{i}_j$ is the set of user $\boldsymbol{i}$'s neighborhood items. The \textbf{minus} sign in Eq.~(\ref{eq:attention}) is used to assign a higher attention score for a lower distance between two items ($j$, $k$).

We note that the $\mathcal{L}2$ distance (or Euclidean distance) satisfies four conditions of a metric \footnote{https://en.wikipedia.org/wiki/Metric\_(mathematics)}. While the crucial triangle inequality property of a metric was shown to provide a better performance compared to the inner product \cite{shrivastava2014asymmetric,ram2012maximum,hsieh2017collaborative} in recommendation domains, to our best of knowledge, most of existing attention designs \cite{vaswani2017attention,luong2015effective,lin2017structured,choi2018fine,seo2016hierarchical,bahdanau2014neural,xu2015show} adopted the inner product for measuring attentive scores. Hence, this proposed attention design is the \textbf{first attempt} to bring metric properties into the attention mechanism.

Similar to \cite{tay2018latent}, we limit the number of considering context items by choosing the user $\boldsymbol{i}$'s $\boldsymbol{s}$ most recently consumed items before target item $\boldsymbol{j}$ as the context items of target item $\boldsymbol{j}$. Here, $\boldsymbol{s}$ can be selected via tuning with a development dataset.
The soft attention vector containing attentive contribution scores of $\boldsymbol{s}$ context items toward the target item $\boldsymbol{j}$ of a user $\boldsymbol{i}$ is given as follows:
%\vspace{-5pt}
\begin{align}
\label{eq:all-attentions}
\boldsymbol{a}_{ij} = \begin{bmatrix}
a_{ij1}, 
\cdots, 
a_{ijs}
\end{bmatrix}^T
\end{align}

\subsubsection{Output Module:}
Given the attentive scores $\boldsymbol{a}_{ij}$ in Eq.(\ref{eq:all-attentions}) and the combined vector $\boldsymbol{p}_{ij} \in \mathbb{R}^d$ of the user embedding $\boldsymbol{u}_i^{(o)}$ and item embedding $\boldsymbol{v}_j^{(o)}$ from the output memory $\boldsymbol{U}^{(o)}$ and $\boldsymbol{V}^{(o)}$,
the goal of this output module is to measure a total output distance $\boldsymbol{o}_{ij}^{(SDM)}$ between the output target item embeddings $\boldsymbol{v}_j^{(o)}$ and all the user $\boldsymbol{i}$ 's output context item embeddings $\boldsymbol{v}_k^{(o)} (k \in T_j^i)$ using attention weights $\boldsymbol{a}_{ij}$ and the output query $\boldsymbol{p}_{ij}$ as follows:
%\vspace{-5pt}
\begin{align}
\label{eq:SDM-finaloutput}
o_{ij}^{(SDM)} = \boldsymbol{w}_e^{\top} \boldsymbol{e}_{ij} + b_e
\end{align}
where $\boldsymbol{e}_{ij} \in \mathbb{R}^{d}$ is calculated as follows:
\vspace{-8pt}
\begin{align}
\label{eq:SDM-out-dist-vec}
\boldsymbol{e}_{ij} = \sum_{k \in \mathcal{T}^{i}_j}
\boldsymbol{a}_{ijk} \times square
\Big(
f 	\Big(\mathbf{W}_d
\begin{bmatrix}
\boldsymbol{p}_{ij} \\
\boldsymbol{v}_k^{(o)}
\end{bmatrix} + \boldsymbol{b}_d
\Big)
\Big)
\vspace{-5pt}
\end{align}
In here, let $\boldsymbol{r}_{ijk} = f \Big(\mathbf{W}_d \begin{bmatrix} \boldsymbol{p}_{ij} \\ \boldsymbol{v}_k^{(o)} \end{bmatrix} + \boldsymbol{b}_d \Big)$.  Similar to the previously discussed intuition in Eq~(\ref{eq:pre-attention}), $\boldsymbol{r}_{ijk}$ is a flexible combination between $\boldsymbol{p}_{ij}$ and each output context item embeddings $\boldsymbol{v}_k^{(o)}$; $square(\cdot)$ is an element-wise squared function.
Our idea in Eq.~(\ref{eq:SDM-finaloutput}), (\ref{eq:SDM-out-dist-vec}) is similar to the idea in Eq.~(\ref{eq:gmp-layerl}), (\ref{eq:gmp-layero}) of the SDP model. First, in Eq.~(\ref{eq:SDM-out-dist-vec}), each context item $k$ will attentively contribute to the target item $j$ via a squared Euclidean measure. Second, in Eq.~(\ref{eq:SDM-finaloutput}), each non-negative dimension in $e_{ij}$ will be considered as a distance dimension and we use an edge-weights layer $\boldsymbol{w}_e$ to combine them flexibly. When there is only one context item in $\mathcal{T}_j^i$, then in Eq.~(\ref{eq:SDM-out-dist-vec}), the attention score $\boldsymbol{a}_{ijk}$=1.0, leading to $\boldsymbol{e}_{ij} = square(\boldsymbol{r}_{ijk})$, which is similar to Eq.~(\ref{eq:gmp-layerl}). In this case, SDM will measure the distance between target item $j$ and context item $k$ in the same way as SDP model does. Note that Eq.~(\ref{eq:SDM-out-dist-vec}) is similar to Eq.~(\ref{eq:gmp-layero}) so SDM can also learn a signed distance value, which also provides a more fine-grained distance compared to a general distance value.

%	where $a_{ijk}$ is the soft attention weights as shown in Eq.(~\ref{eq:attention}), and given that
%	\vspace{-5pt}
%	\begin{align}
%	\label{eq:MMN_output_module}
%	\boldsymbol{p}_{ij} = \mathbf{W}_d
%	\begin{bmatrix}
%	\boldsymbol{u}_i^{(o)}\\
%	\boldsymbol{v}_j^{(o)}
%	\end{bmatrix} + \boldsymbol{b}_d
%	\end{align}

\begin{figure}
	\centering
	\includegraphics[width=0.45\textwidth]{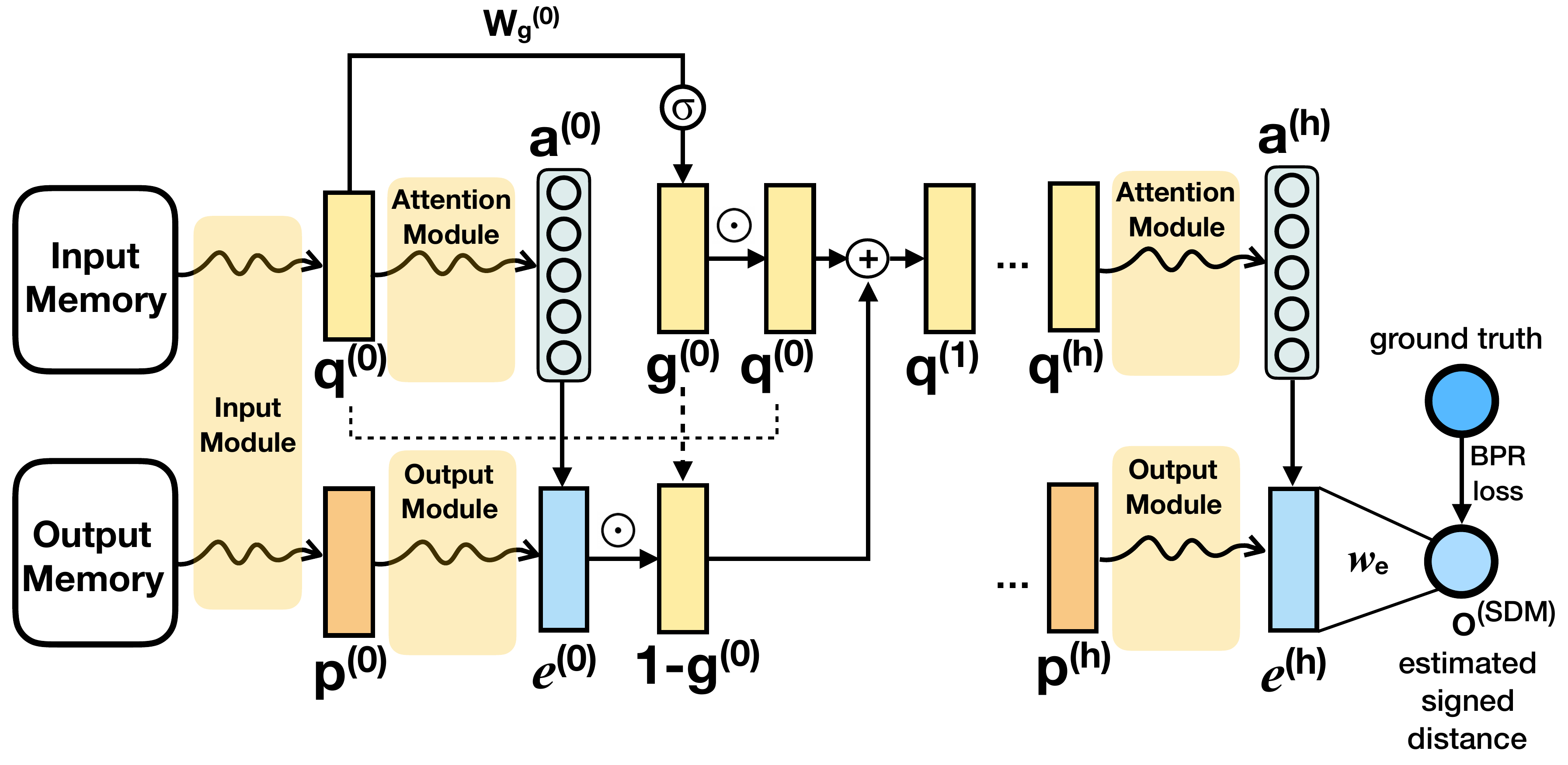}
	\vspace{-10pt}
	\caption{
		The illustration of our multi-hop SDM.
	}
	\label{fig:MMN-multihop}
	\vspace{-10pt}
\end{figure}

\subsubsection{Multi-hop SDM:}
Inspired by previous work \cite{sukhbaatar2015end} where the multi-hop design helped to refine the attention module in Memory Network, we also integrate multiple hops to further extend our SDM model to build a deeper network (Figure~\ref{fig:MMN-multihop}).
As the gated multi-hop design \cite{liu2017gated} was shown to perform better than the original multi-hop design with a simple residual connection in \cite{sukhbaatar2015end}, we employ this gated memory update from hop to hop as follows:
\vspace{-5pt}
\begin{align}
\boldsymbol{g}^{(h-1)} &= \sigma(\mathbf{W}_{g}^{(h-1)} \boldsymbol{q}^{(h-1)} + \boldsymbol{b}_g^{(h-1)}) \\
\boldsymbol{q}^{(h)} &= (1-\boldsymbol{g}^{(h-1)}) \odot \boldsymbol{e}^{(h-1)} + \boldsymbol{g}^{(h-1)} \odot \boldsymbol{q}^{(h-1)}
\end{align}
where $\boldsymbol{q}^{(h-1)}$ is the input query embedding as shown in Eq.~(\ref{eq:MMN_input_module}) at hop $h-1$, $\mathbf{W}_{g}^{(h-1)}$ and bias $\boldsymbol{b}_g^{(h-1)}$ are hop-specific parameters, $\sigma$ is the sigmoid function, ${e}^{(h-1)}$ is the output of Eq.~(\ref{eq:SDM-out-dist-vec}) at hop $h-1$, $\boldsymbol{q}^{(h)}$ is the input query embedding at the next hop $h$.
So the attention could be updated at hop $h$ accordingly using $\boldsymbol{q}^{(t)}$ as follows:
%\begin{align}
%\alpha_{ijk}^{(t)} = \text{exp}\Big(-\Big\Vert{f\Big(\mathbf{W}_b
%    \begin{bmatrix}
%    \boldsymbol{q}_{ij}^{(t)} \\
%    \boldsymbol{v}_k^{(i)}
%    \end {bmatrix} + \boldsymbol{b}_b\Big)}\Big\Vert_2\Big)
%\end{align}
\vspace{-5pt}
\begin{align}
\alpha_{ijk}^{(h)} =  \frac{exp(-z^{(h)}_{ijk})}{\sum_{p \in \mathcal{T}^{i}_j} exp(-z^{(h)}_{ijp})}
\text{, where} \;
z_{ijk}^{(h)} =  \Big\Vert {f \Big(\mathbf{W}_c^{(h)}
	\begin{bmatrix}
	\boldsymbol{q}^{(h)}_{ij} \\
	\boldsymbol{v}_k^{(i)}
	\end {bmatrix} +
	\boldsymbol{b}_c
	\Big)}
\Big\Vert _2 ^2
\end{align}
\vspace{-5pt}
%where $z^{(h)}_{ijk}$ is measured by:
%\begin{align}
%\label{eq:pre-attention-multi-hops}
%z_{ijk}^{(h)} =  \Big\Vert {f \Big(\mathbf{W}_c^{(h)}
%	\begin{bmatrix}
%	\boldsymbol{q}^{(h)}_{ij} \\
%	\boldsymbol{v}_k^{(i)}
%	\end {bmatrix} +
%	\boldsymbol{b}_c
%	\Big)}
%\Big\Vert _2 ^2
%\end{align}
%\vspace{-5pt}

The multi-hop architecture with gated design further refines the attention for different users based on the previous output from hop to hop.
Hence, if the final hop is $h$ then the SDM model with $h$ hops, denoted as \emph{SDM-h}, will use $\boldsymbol{a}_{ij}^{(h)}$ to yield a final signed distance score as follows:
\vspace{-5pt}
\begin{align}
\label{eq:SDM-finaloutput-multi-hops}
o_{ij}^{(SDM-h)} = \boldsymbol{w}_e^{\top} \boldsymbol{e}_{ij}^{(h)} + b_e^{(h)}
\end{align}
where $\boldsymbol{e}_{ij}$ is calculated as:
\vspace{-5pt}
\begin{align}
\label{eq:SDM-out-dist-vec-multi-hops}
\boldsymbol{e}_{ij}^{(h)} = \sum_{k \in \mathcal{T}^{i}_j}
\boldsymbol{a}_{ijk}^{(h)} \times square
\Big( f 	\Big(\mathbf{W}_d^{(h)}
\begin{bmatrix}
\boldsymbol{p}_{ij}^{(h)} \\
\boldsymbol{v}_k^{(o)}
\end{bmatrix} + \boldsymbol{b}_d^{(h)}
\Big)
\Big)
\end{align}
\vspace{-5pt}

%	\vspace{-10pt}
%	\begin{align}
%	\boldsymbol{o}_{ij}^{(T)} = \sum_{k \in \mathcal{T}^{i}_j}  a_{ijk}^{(T)} \Big(\mathbf{W}_c
%	\begin{bmatrix}
%	\boldsymbol{p}_{ij} \\
%	\boldsymbol{v}_k^{(i)}
%	\end{bmatrix} + \boldsymbol{b}_c \Big)
%	\end{align}
%	At last, a final fully connected layer is applied to obtain the distance score,
%	\vspace{-5pt}
%	\begin{align}
%	o_{ij} = \boldsymbol{w}_o^{\top} \boldsymbol{o}_{ij}^{(T)} + b_o
%	\end{align}
%where $\phi(\cdot)$ is the regularization function for the output.
%We choose \textit{absolute function} so SDM could be viewed as the generalization of Minkowski distance, but other functions such as $\text{exp}(\cdot)$ could also be used.

\noindent\textbf{Weight constraints in multi-hop SDM model:} To save memory, we use the global weight constraint in multi-hop SDM. Particularly, input memory $\boldsymbol{U^{(i)}}, \boldsymbol{V^{(i)}}$ and output memory $\boldsymbol{U^{(o)}}, \boldsymbol{V^{(o)}}$ are shared among different hops. All the weights are shared from hop to hop $\boldsymbol{W}_a^{(1)}$ = $\boldsymbol{W}_a^{(2)}$ = ... = $\boldsymbol{W}_a^{(h)}$; $\boldsymbol{W}_b^{(1)}$ = $\boldsymbol{W}_b^{(2)}$ = ... = $\boldsymbol{W}_b^{(h)}$; $\boldsymbol{W}_c^{(1)}$ = $\boldsymbol{W}_c^{(2)}$ = ... = $\boldsymbol{W}_c^{(h)}$;  $\boldsymbol{W}_d^{(1)}$ = $\boldsymbol{W}_d^{(2)}$ = ... = $\boldsymbol{W}_d^{(h)}$; and so do all bias terms. The gate weights are also global weights: $\boldsymbol{W}_g^{(1)}$ = $\boldsymbol{W}_g^{(2)}$ = ... = $\boldsymbol{W}_g^{(h)}$.

\subsection{Signed Distance-based Deep Memory Recommender (SDMR)}
Now we propose Signed Distance-based Deep Memory Recommender (SDMR), a hybrid network that combines SDP and SDM.
The first approach to combine them is to employ a weighted summation of the output scores from SDP and SDM as follows:
\vspace{-2pt}
\begin{align}
\label{eq:sdmr-trivial-combination}
o = \beta o^{\text{(SDP)}} + (1-\beta) o^{\text{(SDM)}}
\end{align}
where $o^{\text{(SDP)}}$ is the signed distance score obtained at Eq.~(\ref{eq:gmp-layero}), $o^{\text{(SDM)}}$ is the signed distance score obtained at Eq.~(\ref{eq:SDM-finaloutput-multi-hops}), and $\beta \in [0, 1]$ is a hyper-parameter to control the contribution of SDP and SDM. When $\beta$=0, SDMR becomes SDM. When $\beta$=1, SDMR becomes SDP.

However, to avoid tuning an additional hyper-parameter $\beta$, we do not use Eq.~(\ref{eq:sdmr-trivial-combination}) for SDMR. Instead, we let SDMR self-learns the combination of SDM and SDM as follows:
\vspace{-2pt}
\begin{align}
\label{eq:sdmr-distance}
o = ReLU \bigg( \boldsymbol{w}_{u}^{\top} \begin{bmatrix} \boldsymbol{e}^{(\ell+1)} \\ \boldsymbol{e}^{(h)}  \end{bmatrix} + b_{u} \bigg)
\end{align}
where $\boldsymbol{e}^{(\ell+1)}$ is the final layer embedding from SDP and is obtained at Eq.~(\ref{eq:gmp-layerl}), $\boldsymbol{e}^{(h)}$ is the final hop output from the multi-hop SDM obtained at Eq.~(\ref{eq:SDM-out-dist-vec-multi-hops}). We note that SDP and SDM are first pre-trained separately using the BPR loss function (see the next section). Then, we obtain $\boldsymbol{e}^{(\ell+1)}$ from SDP, and $\boldsymbol{e}^{(h)}$ from SDM, and keep them fixed in Eq.~(\ref{eq:sdmr-distance}) to learn $\boldsymbol{w}_{u}$ and $b_{u}$. We use \texttt{ReLU} in Eq.~(\ref{eq:sdmr-distance}) because \texttt{ReLU} encourages sparse activations and helps to reduce over-fitting when combining the two components SDP and SDM.
%In another word, SDMR just learns the combination weights $w_u$ and bias $b_u$ between the two components.

%\subsection{A modification of SDMR for non-personalized recommendation task:}
\subsection{Loss Functions}
We adopt the Bayesian Personalized Ranking (BPR) as our loss function, which is similar to the idea of AUC (area under the curve):
\vspace{-5pt}
\begin{equation}
\label{eq:BPR}
\noindent
\resizebox{0.42\textwidth}{!}{$
	\begin{aligned}
	\mc{L} = \operatorname*{argmin}_\theta \Big( -\sum_{(u, i^+, i^-)} \text{log } \sigma(o_{ui^-} - o_{ui^+}) + \lambda \Vert \theta \Vert^2 \Big)\\
	%& \text{and } \sigma(x) = \frac{1}{1+ e^{-x}}
	\end{aligned}
	$}
%\vspace{-2pt}
\end{equation}
where we uniformly sample tuples in a form of $(u, i^+, i^-)$ for user $u$ with positive item (consumed) $i^+$ and negative item (unconsumed) $i^-$. $\lambda$ is a hyper-parameter to control the regularization term, and $\sigma(\cdot)$ is the sigmoid function. Note that other pairwise probability functions could be plugged in Eq.~(\ref{eq:BPR}) to replace $\sigma(\cdot)$.
Both SDP and SDM are end-to-end differentiable since we uses soft attention over the output memory.
Hence, we can utilize back-propagation to learn our models with stochastic gradient descent or Adam \cite{kingma2014adam}.

\section{Empirical Study}
%In this section, we evaluate our proposed SDP, SDM, and SDMR 
We evaluate our SDP, SDM, SDMR 
models against ten state-of-the-art baselines in two recommendation tasks: (i) \emph{general recommendation task}, and (ii) \emph{shopping basket-based recommendation task}. We mainly aim to answer the following research questions (RQs):
%\begin{itemize}
\squishlist
	\item \textbf{RQ1:} How do SDP, SDM, and SDMR perform compared to other state-of-the-art models in both general recommendation task and shopping basket-based recommendation task?
	%\item \textbf{RQ2:} How do SDP, SDM and SDMR perform compared to other state-of-the-art models when (i) varying top-k recommendation and (ii) varying embedding size?
	\item \textbf{RQ2:} Why/How does the multi-hop design help to improve the proposed models' performance?
%	\item \textbf{RQ4:} What are the memory expense and runtime of SDP, SDM, and SDMR compared to other baselines?
	%\item \textbf{RQ5:} How does multi-hop design visually help for recommendation tasks?
\squishend	
%\end{itemize}
\subsection{Datasets}
%In each recommendation task, we conduct experiments on the following datasets:

\noindent\textbf{General recommendation task:} In this task, we evaluate our proposed models and state-of-the-art methods using different datasets with various density levels as follows:
\squishlist
%\begin{itemize}
	\item \textbf{Movielens} \cite{resnick1994grouplens}: It is a widely adopted benchmark dataset for collaborative filtering evaluation. We use two versions of this benchmark dataset, namely MovieLens100k (or ML-100k) and MovieLens1M (or ML-1M).
	\item \textbf{Netflix Prize} \footnote{https://www.netflixprize.com/}: It is a real-world dataset collected by Netflix. This dataset was collected from 1999 to 2005, and consists of 463,435 users and 17,769 items with 56.9M of interactions. Since the dataset is extremely large, we subsample the Netflix dataset by randomly picking one-month data for evaluation.
	\item \textbf{Epinions} \cite{massa2007trust} \footnote{http://www.trustlet.org/downloaded\_epinions.html}: It is an online rating dataset where users can share product feedback by giving explicit ratings and reviews.
%\end{itemize}
\squishend
In preprocessing preparation, we adopted a popular k-core preprocessing step \cite{he2016ups,liang2018variational,tran2018regularizing} (with \emph{k-core} = 5) to filter out inactive users with less than five ratings and items which are consumed by less than five users. Since ML-100k and ML-1M are already preprocessed, we only apply 5-core preprocessing step on the Netflix and Epinions datasets.
%This results in 1,888 users and 3,724 items.
We also binarize the rating scores as implicit feedback by converting all observed rating scores as positive interactions and the remaining as negative interactions. The statistics of the four datasets are summarized in Table \ref{table:datasets}.

\noindent\textbf{Shopping basket-based recommendation task:} 
We use two real-world transaction datasets as follows:
%We evaluate our proposed models on two real-world transaction datasets as follows:
\squishlist
%\begin{itemize}
	\item \textbf{IJCAI-15} \footnote{https://tianchi.aliyun.com/datalab/dataSet.htm?id=1}: %It is a well-known shopping basket-based dataset. 
	It consists of shopping logs of users from Tmall \footnote{https://www.tmall.com}. Since the original dataset is extremely large scale. We subsample IJCAI-15 by randomly picking 20k transactions for evaluation.
	\item \textbf{Tafeng} \footnote{http://stackoverflow.com/questions/25014904/download-link-for-ta-feng-grocery-dataset}: It is a grocery store transaction data. It contains four month transaction data from November 2000 to February 2001 by T-Feng supermarket.
%\end{itemize}
\squishend

Users in both IJCAI-15 and Tafeng datasets are logged under four types of actions: \emph{click}, \emph{add-to-cart}, \emph{purchase}, and \emph{add-to-favourite}. We consider all the four types as the \emph{click} action. We only keep transactions with at least five items. This is because we will take one item out for testing, another item for development. In the remaining three items, one will be taken out as a target item and the two items will be used as the context items. Attentive scores will be assigned to the context items. In each of original transactions, we generate data instances of the format $<\mathbf{c}, v_c>$ where $v_c$ is the target/predicting item and $\mathbf{c}$ is a set of all other items in the same transaction with $v_c$. %(i.e. $\mathbf{c}$ can be called as context items or context items of $v_c$).
In particular, in each transaction $t$, each time we pick one item out as a target item and leave the rest of items in $t$ as corresponding context items. Subsequently, for each transaction $t$ containing $|t|$ items, we can generate $|t|$ data instances. The statistics of the two transactional datasets are summarized in Table~\ref{table:transaction-datasets}.

For an easy reference, we call (ML-100k, ML-1M, Netflix, Epinions) as \emph{Group-1 dataset} and (IJCAI-15, Ta-Feng) as \emph{Group-2 datasets}.
%--Dataset Table
\begin{table}[t]
	\centering
	\tiny
	\caption{Statistics of the four datasets in the general recommendation task.}
	\vspace{-5pt}
	\label{table:datasets}
	\resizebox{1.0\linewidth}{!}{
		\begin{tabular}{lcccc}
			\toprule
			Statistics         & ML-100k   & ML-1M     & Netflix    & Epinions \\
			\midrule
			\# of users        & 943       & 6,040     & 1,888      & 23,137\\
			\# of items        & 1,682     & 3,706     & 3,724      & 23,585\\
			\# of interactions & 100,000   & 1,000,209 & 103,254    & 461,982\\
			Density (\%)       & 6.3\%     & 4.5\%     & 1.5\%      & 0.08\%\\
			\bottomrule
		\end{tabular}
	}
	\vspace{-10pt}
\end{table}

\begin{table}[t]
	\centering
	\tiny
	\caption{Statistics of the two real-world transactional datasets in the shopping basket-based recommendation task.}
	\vspace{-10pt}
	\label{table:transaction-datasets}
	\resizebox{0.86\linewidth}{!}{
		\begin{tabular}{lcc}
			\toprule
			Statistics		                      	& IJCAI-15   & Tafeng  \\
			\midrule	
			\# of users 	                    	& 2,433      & 22,851   \\
			\# of items     	                	& 4,534      & 22,291   \\
			%max \#items in a transaction     &            &          \\
			avg \# of items in a transaction     	& 6.28       &  9.28    \\
			%min \#items in a transaction     &            &          \\
			\# of generated instances       		& 15,422     & 523,653  \\
			Density (\%)                    		& 0.14\%     & 0.10\%   \\
			\bottomrule
		\end{tabular}
	}
	\vspace{-10pt}
\end{table}
%--

\subsection{Baselines and State-of-the-art Methods}
We compared our proposed models against several strong baselines in the general recommendation task as follows:
\squishlist
%\begin{itemize}
	\item \textbf{Item{\sc{knn}}} \cite{sarwar2001item}: It is an item neighborhood-based collaborative filtering method. It exploited cosine item-item similarities to produce recommendation results.
	\item \textbf{Bayesian Personalized Ranking (MF-BPR)} \cite{rendle2009bpr}: It is a state-of-the-art pairwise matrix factorization method for implicit feedback datasets. It minimizes $\sum_{i}\sum_{j^{+},j^{-}} -log\sigma(u_i^Tv_{j^+}$ - $u_i^Tv_{j^-})$ + $\lambda(||u_i||^2 + ||v_{j^+}||^2)$ where ($u_i$, $v_{j^+}$) is a positive interaction and ($u_i$, $v_{j^-}$) is a negative sample.
	\item \textbf{Sparse LInear Method ({\sc{slim}})} \cite{ning2011slim}: %It is a sparse linear model which
	It learns a sparse item-item similarity matrix by minimizing the squared loss $||A-AW||^2 + \lambda_1||W|| + \lambda_2||W||^2$, where A is a $m\times n$ user-item interaction matrix and W is a $n \times n$ sparse matrix of aggregation coefficients of context items.
	\item \textbf{Collaborative Metric Learning (CML)} \cite{hsieh2017collaborative}: It is a state-of-the-art collaborative metric-based model that utilizes Euclidean distance to measure similarities between users and items. %Its intuition is to pull users and positive items closer and push away negative items.
	%For fair comparison, we learn CML with BPR loss by minimizing $-\sum_{i, j^+, j^-}log (\sigma(d(u_i, v_{j^-})^2 - d(u_i, v_{j^+})^2))$, where $d(u_i, v_{j^+})^2$ is a squared Euclidean distance of a positive interaction ($u_i$, $v_{j^+}$) and  $d(u_i, v_{j^-})^2$  is a squared Euclidean distance of a negative sample ($u_i$, $v_{j^-}$).
	For fair comparison, we learn CML with BPR loss by minimizing $-\sum_{i, j^+, j^-}log (\sigma(||u_i - v_{j^-}||_2^2 - ||u_i - v_{j^+}||_2^2))$, where $||\cdot||_2^2$ is a squared Euclidean distance, ($u_i$, $v_{j^+}$) is a positive interaction and $(u_i, v_{j^-})$  is a negative sample.
	
	\item \textbf{Neural Collaborative Filtering (NeuMF++)} \cite{he2017neural}: It is a state-of-the-art matrix factorization method using deep learning architecture. We use a pre-trained NeuMF to achieve its best performance, and denote it as NeuMF++.
	\item \textbf{Collaborative Memory Network (CMN++)} \cite{ebesu2018collaborative}:  It is a state-of-the-art memory network based recommender. Its architecture follows traditional user neighborhood based collaborative filtering approaches. It adopts a memory network to assign attentive weights for other similar users.
	%\item Generalized Metric-based Perceptron (SDP) (this paper): our proposed model. It measures explicitly the distance between the target user and the target item.
	%\item Metric-based Memory Model (SDM) (this paper): our proposed model. It measures the attentive summation of personalized distances between the target item and the user's consumed items.
	%\item Deep Metric-based Memory Recommender (SDMR) (this paper): our proposed model that combines SDP and SDM.
%\end{itemize}
\squishend
%We exclude baselines that worked worse than NeuMF such as: Factorization Machines \cite{rendle2010factorization} and eALS \cite{he2016fast}.
Even though our approaches do not model the order of consumed items in the user's purchase history (e.g. rigid orders of items), since we consider latest $s$ items as the context items to predict the next item, we still compare our models with some key sequential models to further show our models' effectiveness as follows:
\squishlist
%\begin{itemize}
	\item \textbf{Personalized Ranking Metric Embedding (PRME)} \cite{feng2015personalized}: \\ Given a user $u$, a target item $j$, and a previous consumed item $k$, it models a personalized first-order Markov behavior with two components: $d_{ujk} = \alpha||v_u - v_j||^2 + (1-\alpha)||v_k - v_j||^2$, 
	%where $||v_u - v_j||^2$ is a squared $\mathcal{L}2$ distance of ($u,j$), and $||v_k - v_j||^2$ is a squared $\mathcal{L}2$ distance of ($k,j$). 
	where $||\cdot||_2^2$ is a squared $\mathcal{L}2$ distance. 
	Then PRME is learned by minimizing BPR loss.
	\item \textbf{PRME\_s:} It is our extension of PRME, where the distance between the target item $j$ and the previous consumed item $k$ is replaced by the average distance between $j$ and each of previous $s$ items: $d_{ujs} = \alpha||v_u - v_j||^2 + (1-\alpha)\frac{1}{|s|}\sum_{k \in s}||v_k - v_j||^2$. We use BPR loss to learn PRME\_s.
	\item \textbf{Translation-based Recommendation (TransRec)} \cite{he2017translation}: It uses first-order Markov and considers a user $u$ as a translator of his/her previous consumed item $k$ to a next item $j$. In another word, $prob(j|u, k) \propto \beta_j - d(u + v_k - v_j)$ where $\beta_j$ is an item bias term, $d$ is a distance function (e.g. $\mathcal{L}1$ or $\mathcal{L}2$ distance). We use $\mathcal{L}2$ distance because it was shown to perform better than $\mathcal{L}1$ \cite{he2017translation}. TransRec is then learned with BPR loss.
	\item \textbf{Convolutional Sequence Embedding Recommendation\\(Caser)} \cite{tang2018personalized}: It is a state-of-the-art sequential model. It uses convolution neural network with many horizontal and vertical kernels to capture the complex relationships among items.
%\end{itemize}
\squishend
%The strong sequential baselines above surpassed many other sequential models such as: TransRec outperformed Factorized Markov Chain (FMC)\cite{rendle2010factorizing} Factorized Personalized Markov Chain (FPMC) \cite{rendle2010factorizing}, Hierarchical Representation Model (HRM) \cite{wang2015learning}, Caser outperformed GRU4Rec \cite{hidasi2015session} and Fossil \cite{he2016fusing}, so we exclude them in our evaluation.
The strong sequential baselines above surpassed many other sequential models such as: TransRec outperformed FMC\cite{rendle2010factorizing}, FPMC \cite{rendle2010factorizing}, HRM \cite{wang2015learning}; Caser surpassed GRU4Rec \cite{hidasi2015session} and Fossil \cite{he2016fusing}, so we exclude them in our evaluation.

\noindent\textbf{Comparison:} In the general recommendation task, we compare our proposed models with all \textbf{ten} strong baselines listed above. In the shopping basket-based recommendation task, since the sequential models often work better than general recommendation-based models (see Table~\ref{table:PerformanceComparison}), we only compared our proposed models with sequential baselines. We name general recommendation baselines (i.e. ItemKNN, BPR, SLIM, CML, NeuMF++, CMN++) as \emph{Group-1 baselines}, and call sequential baselines (i.e. PRME, PRME\_s, TransRec, Caser) as \emph{Group-2 baselines} for an easy reference.

%We exclude models that utilized additional information (such as reviews, etc.) since we do not use contextual information in our models. We also do not compare our models with temporal, and sequential models since we did not model the order of consumed items in our proposed models.

\subsection{Experimental Settings}
%To evaluate our proposed models and the baselines,
\noindent\textbf{Protocol:} We adopt the widely used \emph{leave-one-out} setting \cite{he2017neural,xue2017deep}, in which for each user, we reserve her last interaction as the test sample. If there are no timestamps available in the dataset, then the test sample is randomly drawn.
Among the remaining data, we randomly hold one interaction for each user to form the development set, while all others are utilized as the training set.
Since it is very time-consuming and unnecessary to rank all the unobserved items for each user, we follow the standard strategy to randomly sample 100 unobserved items for each user. Then, we rank them together with the test item \cite{he2017neural,koren2008factorization}.

\noindent\textbf{Assigning item orders:} Sequential models need rigid orders of consumed items but consumed items in the same transaction (in IJCAI-15 and TaFeng datasets) are assigned the same timestamp of the transaction containing these items. Hence, we assigned the item timestamps where the orders of items are kept as in the original dataset. This may give credits to sequential models but not our methods (because our methods will use all consumed items in the same transaction as context items and do not model the item orders).

\noindent\textbf{Hyper-parameters selection:}
%The hyper-parameter of all models are tuned using the development set.
We perform a grid search for the embedding size $d$ from $\{8, 16, 32, 64, 128\}$ and regularization terms from $\{0.1, 0.01, 0.001, 0.0001, 0.00001\}$ in all the models. We select the best number of hops for CMN++ and our SDM from $\{1, 2, 3, 4\}$. In NeuMF++, we select the best number of MLP layers from $\{1, 2, 3\}$.
In our models, we fix the batch size to $256$. We adopt Adam optimizer \cite{kingma2014adam} with a fixed learning rate of 0.001. Similar to CMN++ and NeuMF++, the number of negative samples is set to 4.
We use one layer perceptron for SDP (more complex datasets may need more than one layer to get better results). 
%We initialize the user and item embeddings using $\mathcal{N}(\mu=0, \sigma=\frac{1}{d})$, and initialize the edge-weights layers using \emph{He normal initializer} (e.g. $\boldsymbol{w^{(o)}}$, $\boldsymbol{w}_e$, $\boldsymbol{w}_u$ in Eq.~(\ref{eq:gmp-layero}), (\ref{eq:SDM-finaloutput-multi-hops}), (\ref{eq:sdmr-distance}), respectively).
%(\ie the last layer in SDP or SDM that maps a input vector to a final distance value)
%For our proposed SDM model, we use \emph{tanh} activation and perform a grid search of the memory hop from [1, 2, 3, 4].
%We fix \emph{identity} activation for SDP model.
%The selection of activation functions in our proposed methods will be discussed in a later section.
In the four datasets used in general recommendation task (e.g ML-100k, ML-1M, Netflix, Epinions), to avoid too many \emph{zero paddings} for users with a smaller number of consumed items or too many context items are kept in the memory, which unnecessarily slow down the model's execution, we follow \cite{tay2018latent} to limit the number of context items using latest \emph{s} consumed items. We search s in \{5, 10, 20\}. In the two shopping basket-based recommendation datasets (i.e. IJCAI-15 and TaFeng), since the maximum number of items in a transaction is small (e.g. 13 in IJCAI-15, and 18 in TaFeng), we consider all the other items in the same transaction with the target item as its context items.
All the hyper-parameters are tuned using the development dataset.
Our source code is available at: \emph{https://github.com/thanhdtran/SDMR}.
%The recent success of transitional/sequential recommenders \cite{rendle2010factorizing,feng2015personalized,wang2015learning} suggested that we do not need to encode all user's historical items which can cause unnecessary \emph{zero paddings} for users with a smaller number of consumed items.
%This will also unnecessarily slow down the model's execution.
%Therefore, we consider a user's latest \emph{s} items to predict her next item. We fix $s=5$ in all experiments.

\noindent\textbf{Evaluation Metrics:}
We evaluate all models' performance by two widely used metrics: Hit Ratio (\emph{hit}@$k$), and Normalized Discounted Cumulative Gain (NDCG@$k$), where $k$ is a truncated number or \emph{top-k} item recommendation.
Intuitively, \emph{hit}@$k$ shows whether the test item is in the \emph{top-k} list or not, while \emph{NDCG}@$k$ accounts for the position of the hits by assigning higher scores to the hits at top ranks and downgrading the scores to hits by $log_2$ at lower ranks.

%--overall performance
\label{sec:exp}
\begin{table*}[t]
	\centering
	\tiny
	\caption{General Recommendation Task: Overall performance of the baselines, and our proposed SDP, SDM, and SDMR on four datasets. The last four lines show the relative improvement of the SDM and SDMR over the best baseline method in General Recommenders (Group 1) and Sequential Recommenders (Group 2), respectively.}
	\vspace{-5pt}
	\label{table:PerformanceComparison}
	\resizebox{0.95\textwidth}{!}{
		\begin{tabular}{llccccccccc}
			\toprule
			\multirow{2}{*}{\text{Method type}} & \multirow{2}{*}{\text{Method}} & \multicolumn{2}{c}{\textbf{ML-100k}} &\multicolumn{2}{c}    {\textbf{ML-1M}} & \multicolumn{2}{c}{\textbf{Netflix}} & \multicolumn{2}{c}{\textbf{Epinions}}\\
			\cmidrule{3-10}
			& & \textit{hit}@10 & NDCG@10 & \textit{hit}@10 & NDCG@10 & \textit{hit}@10 & NDCG@10 & \textit{hit}@10 & NDCG@10
			\\
			\midrule
			\multirow{6}{*}{\pbox{1cm}{General\\ Recommenders \\ (Group 1)}}	&
			Item-KNN      & 0.166 & 0.073 & 0.235 & 0.110 & 0.039 & 0.019 & 0.121 & 0.096 \\
			& SLIM        & 0.520 & 0.298 & 0.677 & 0.420 & 0.358 & 0.212 & 0.249 & 0.189 \\
			& MF-BPR      & 0.554 & 0.316 & 0.595 & 0.352 & 0.352 & 0.193 & 0.384 & 0.232\\
			& CML         & 0.596 & 0.326 & 0.662 & 0.390 & 0.447 & 0.254 & 0.376 & 0.237 \\
			& NeuMF++     & 0.623 & 0.341 & 0.716 & 0.438 & 0.509 & 0.279 & 0.428 & 0.274 \\
			& CMN++       & 0.620 & 0.344 & 0.729 & 0.442 & 0.523 & 0.293 & 0.423 & 0.272 \\
			\midrule
			\multirow{4}{*}{\pbox{1cm}{Sequential\\ Recommenders \\ (Group 2)}}
			& PRME        & 0.638 & 0.381 & 0.724 & 0.486 & 0.509 & 0.329 & 0.538 & 0.346 \\
			& PRME\_s     & 0.674 & 0.398 & 0.734 & 0.491 & 0.539 & 0.348 & 0.380 & 0.244 \\
			& TransRec    & 0.684 & 0.402 & 0.770 & 0.524 & 0.511 & 0.345 & 0.551 & 0.357 \\
			& Caser       & 0.674 & 0.386 & \textbf{0.826} & 0.606 & 0.480 & 0.253 & 0.326 & 0.268 \\
			\midrule
			\multirow{3}{*}{\textbf{Ours}}
			& SDP         & 0.616 & 0.349 & 0.694 & 0.424 & 0.497 & 0.279 & 0.416 & 0.266 \\
			& SDM         & \textbf{0.713} & 0.435 & \textbf{0.816} & 0.584 & \textbf{0.584} & 0.379 & \textbf{0.575} & 0.390 \\
			& SDMR        & \textbf{0.695} & \textbf{0.562} & \textbf{0.810} & \textbf{0.662} & \textbf{0.592} &  \textbf{0.449} & \textbf{0.568} & \textbf{0.423} \\
			\midrule
			\multirow{2}{*}{\pbox{1cm}{Compared to \\ Group 1}}
			& Imprv. of SDM  & 14.54\% & 26.51\% & 11.93\% & 32.13\% & 11.71\% & 29.32\% & 34.35\% & 42.34\% \\
			& Imprv. of SDMR & 11.65\% & 63.44\% & 11.11\% & 49.77\% & 13.24\% & 53.20\% & 32.71\% & 54.38\% \\
			\midrule
			\multirow{2}{*}{\pbox{1cm}{Compared to \\ Group 2}}
			& Imprv. of SDM  & 4.24\% &  8.21\% & -1.21\% & -3.63\% &  8.35\% &  8.91\% & 4.36\% &  9.24\% \\
			& Imprv. of SDMR & 1.61\% & 39.80\% & -1.94\% &  9.24\% &  9.83\% & 29.02\% & 3.09\% & 18.49\% \\
			\bottomrule
		\end{tabular}
	}
	\vspace{-10pt}
\end{table*}

\begin{table}[t]
	\centering
	\tiny
	\caption{Shopping basket-based Recommendation Task: Overall performance of the baselines, and our proposed models on two datasets. The last two lines show the relative improvement of the SDM and SDMR over the best baseline.}
	\vspace{-5pt}
	\label{table:ShoppingBasketPerformance}
	\resizebox{0.46\textwidth}{!}{
		\begin{tabular}{lcccc}
			\toprule
			\multirow{2}{*}{Method} & \multicolumn{2}{c}{\textbf{IJCAI-15}}                           & \multicolumn{2}{c}{\textbf{Ta-Feng}}                              \\
			\cmidrule{2-5} & \textit{hit}@10 & NDCG@10 & \textit{hit}@10 & NDCG@10 \\
			\midrule
			PRME            & 0.276     & 0.177     & 0.594     & 0.365                     \\
			PRME\_s         & 0.229     & 0.133     & 0.590     & 0.355                     \\
			TransRec        & 0.262     & 0.168     & 0.622     & 0.401                     \\
			Caser           & 0.173     & 0.096     & 0.605     & 0.373                     \\
			\midrule
			SDP             & 0.323     & 0.201     & 0.633     & 0.401                     \\
			SDM             & 0.316     & 0.189     & \textbf{0.646}     & 0.439                     \\
			SDMR            & \textbf{0.336}     & \textbf{0.222}     & 0.627     & \textbf{0.559}                    \\
			\midrule
			Imprv. of SDM  & 14.49\%   &  6.78\%    & 3.86\%    & 9.48\%                    \\
			Imprv. of SDMR  & 21.74\%   & 25.42\%    & 0.80\%    & 39.40\%                    \\
			\bottomrule
		\end{tabular}
	}
	\vspace{-5pt}
\end{table}
%--end overall performance

\vspace{-5pt}
\subsection{Experimental Results}
\label{subsec:exp-res}
\noindent\textbf{RQ1: Overall results in general recommendation task:}
The performance of our proposed models and the baselines are shown in Table \ref{table:PerformanceComparison}.
First, we observe that SDP significantly outperformed BPR in all four datasets in \emph{Group-1 datasets}, improving \emph{hit}@10 from 8.33$\sim$41.19\%, and {NDCG}@10 from 10.44$\sim$44.56\%.
Although SDP and BPR shared the same loss function, the difference between them is SDP measured a signed distance score between a target user and a target item via a MLP which modeled a non-linear interaction between them, while BPR used Matrix Factorization with inner product. This result confirms the effectiveness of using signed distance based similarity over inner product in the general recommendation task.
Second, we compare SDP with CML. CML worked by trying to minimize the squared Euclidean distance scores between target users and target items.
%CML worked by minimizing a squared Euclidean distance between target users and target items to pull them closer, while maximizing a squared Euclidean distance between target users and negative sampled items to push negative items away from target users.
Our SDP, in another hand, works by minimizing signed distance scores of non-linear interactions (via non-linear activation functions) between target users and target items. We observe that SDP performed better than CML in all \emph{Group-1 datasets}, improving \emph{hit}@10 from 8.33$\sim$11.19\%, and {NDCG}@10 from 7.06$\sim$12.24\%. On average, SDP improved \emph{hit}@10 by 7.5\% and \emph{NDCG}@10 by 9.5\% compared to CML.
Our SDP even gain competitive results compared to NeuMF++ and CMN++. On average, SDP is just slightly worse than NeuMF++ and CMN++ by -2.67\% for \emph{hit}@10, and -1.68\% for \emph{NDCG}@10.
All of these results show the effectiveness of using signed distance in our SDP model.

Next, we compare SDM with neighborhood-based baselines.
Both {\sc{slim}} and item-KNN used previously consumed items of a user to make the prediction for the next item.
SDM significantly outperformed both baselines, improving \emph{hit}@10 from 20.53$\sim$130.92\% and {NDCG}@10 from 39.05$\sim$106.35\% compared with {\sc{slim}}.
It is an obvious result because the neighborhood-based baselines barely measured linear similarities between the target item and the user's consumed items.
In contrast, our SDM produced signed distance scores and assigned personalized metric-based attention weights to each of consumed items that contribute to the target item.

We then compare SDM with CMN++ and NeuMF++.
%First, both SDM and CMN++ exploited memory network for collaborative filtering recommendations. However, we note several key differences between our design and CMN as: (i) our proposed SDM followed item-based neighborhood design, which was shown to perform slightly better than user-based neighborhood design \cite{linden2003amazon,sarwar2001item}; (ii) SDM used our proposed first-attempt personalized metric-based attention mechanism and produces signed distance scores as output, whereas CMN barely exploited a traditional inner product based attention; and (iii) our multi-hop design exploited gated multi-hop design, which was show to perform better than original multi-hop design \cite{liu2017gated}.
SDM outperformed CMN++ in all \emph{Group-1 datasets}, improving \emph{hit}@10 from 11.71$\sim$35.93\% and {NDCG}@10 from 26.51$\sim$43.38\%. On average, it improves \emph{hit}@10 by 18.63\% and \emph{NDCG@10} by 32.84\% compared to CMN++.
This result shows the effectiveness of our personalized metric-based attention with signed distance and item-based neighborhood design over the traditional inner product-based attention in a user-based neighborhood design in CMN++.
SDM also outperformed NeuMF++, improving \emph{hit}@10 from 13.97$\sim$34.35\%, and {NDCG}@10 from 27.42$\sim$42.34\%.
On average, in all \emph{Group-1 datasets}, SDM outperformed all the baselines in ``General Recommenders'' (Group 1), improved \emph{hit}@10 by 18.13\% and \emph{NDCG}@10 by 32.58\% compared to the best baseline in \emph{Group 1}.

Finally, we look at the performance of SDMR model, which is the proposed fusion of SDP and SDM. Compared to SDM, our SDMR insignificantly downgrades SDM on \emph{hit}@10 measurement with a very small amount, but it does help a lot in refining the ranking of items and boosting \emph{NDCG}@10 results. As shown in Table~\ref{table:PerformanceComparison}, SDMR improved from 8.46$\sim$29.20\% for \emph{NDCG}@10, and by 17.37\% for \emph{NDCG}@10 on average compared to SDM in \emph{Group-1 datasets}. SDMR also surpassed all the methods in \emph{Group 1}. On average, SDMR improved \emph{hit}@10 by 17.18\% and {NDCG}@10 by 55.20\% compared to the best model in \emph{Group 1}.

We also compared our models with some strong sequential models in Table~\ref{table:PerformanceComparison}. Sequential models exploited consuming time of items and model their rigid orders, which often lead to a much improved performance compared to general recommendation models in \emph{Group-1 baselines}. As such, compared to the best sequential baseline model, on average, SDM improves \emph{hit}@10 by 3.94\% and \emph{NDCG}@10 by 5.68\% , and SDMR improves \emph{hit}@10 by 3.15\% and \emph{NDCG}@10 by 24.14\% compared to the best sequential model reported in Table~\ref{table:PerformanceComparison}.

\noindent\textbf{Overall results in shopping basket-based recommendation task}: Table~\ref{table:ShoppingBasketPerformance} shows the performance of our models and sequential baselines in \emph{Group-2 datasets}. Again, our models outperformed all the sequential baselines.  
On average, SDM improved \emph{hit}@10 by 9.2\% and \emph{NDCG}@10 by 8.1\%, SDMR improved \emph{hit}@10 by 11.3\% and \emph{NDCG}@10 by 32.4\% compared to the best reported baseline.
\begin{figure}[]
	\centering
	\begin{subfigure}{0.115\textwidth}
		\centering
		\includegraphics[width=\textwidth]{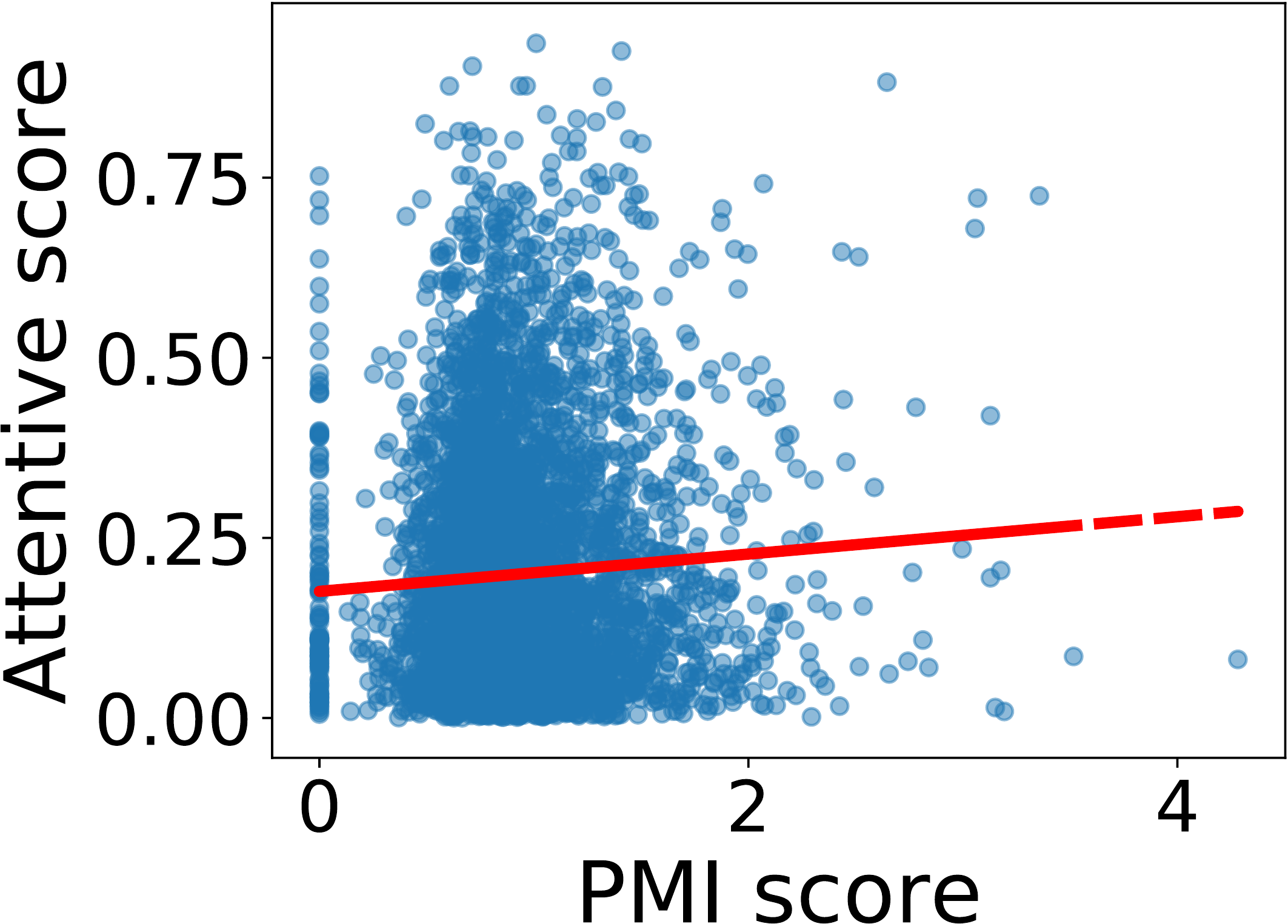}
		\caption{Hop 1.}
		\label{fig:pmi-attention-hop1}
	\end{subfigure}
	\begin{subfigure}{0.115\textwidth}
		\centering
		\includegraphics[width=\textwidth]{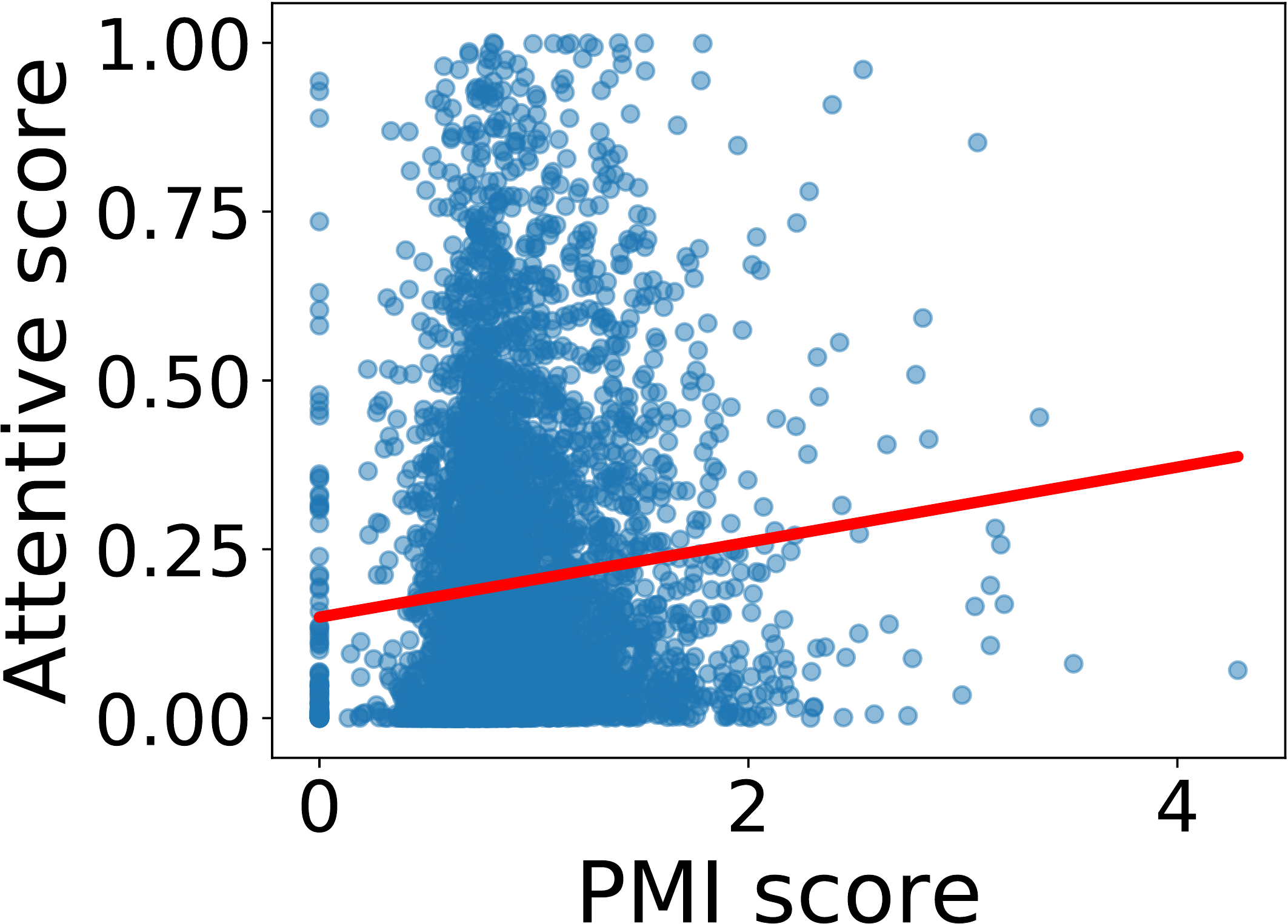}
		\caption{Hop 2.}
		\label{fig:pmi-attention-hop2}
	\end{subfigure}
	\begin{subfigure}{0.115\textwidth}
		\centering
		\includegraphics[width=\textwidth]{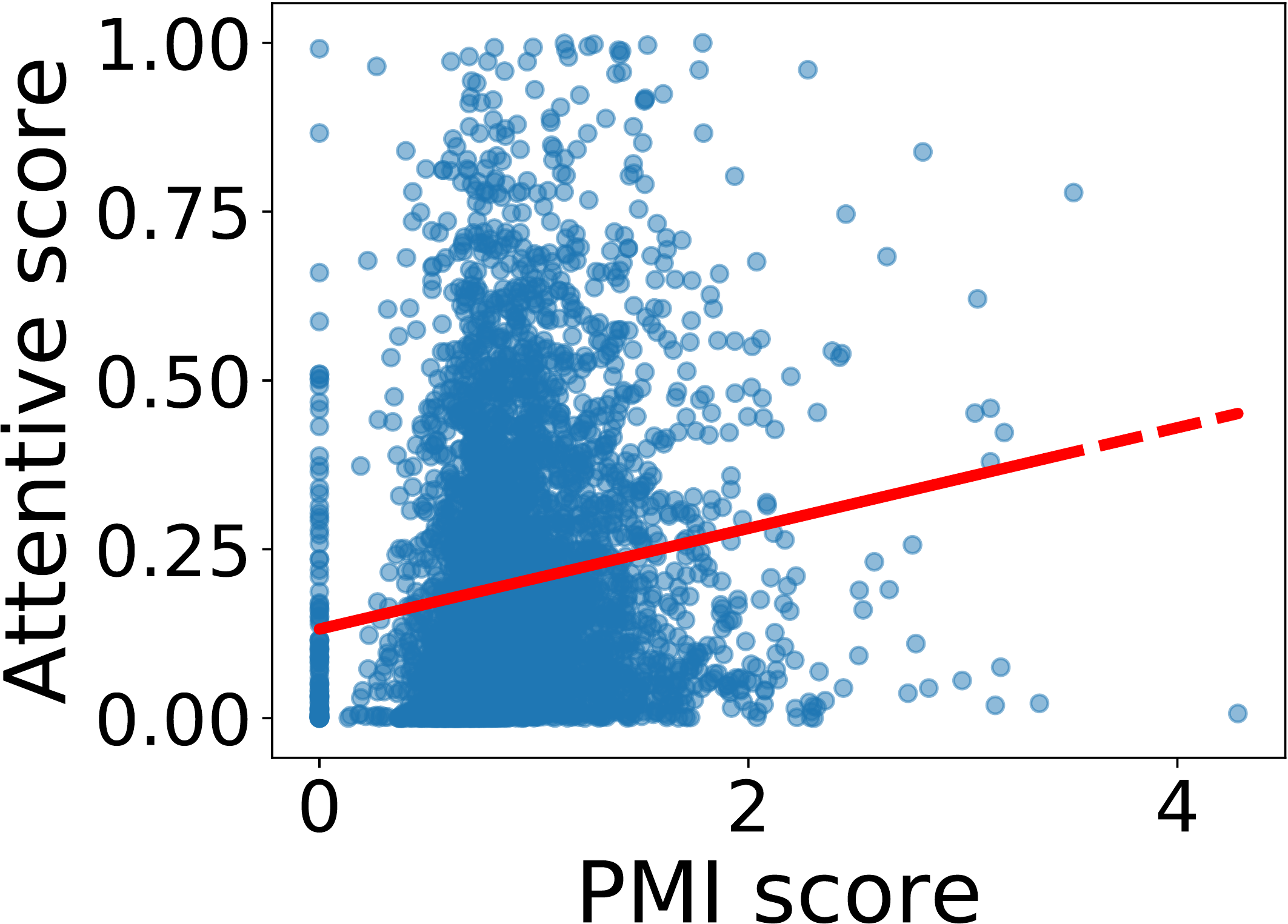}
		\caption{Hop 3.}
		\label{fig:pmi-attention-hop3}
	\end{subfigure}
	\begin{subfigure}{0.115\textwidth}
		\centering
		\includegraphics[width=\textwidth]{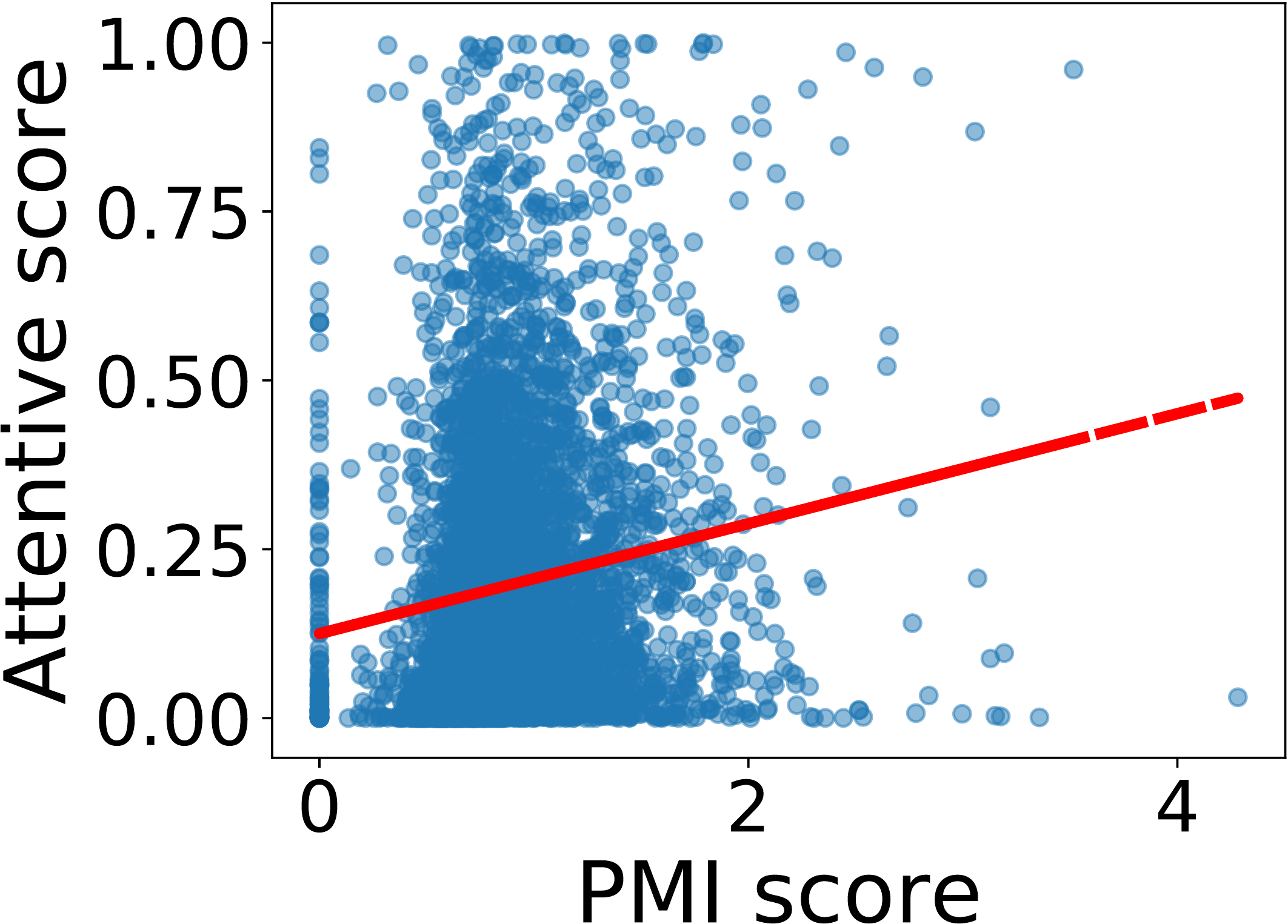}
		\caption{Hop 4.}
		\label{fig:pmi-attention-hop4}
	\end{subfigure}
	\vspace{-10pt}
	\caption{ML-100K: Scatter plots of PMI scores and attentive scores generated by SDM with h hops (h=\{1, 2, 3, 4\} from left to right). The red lines are the linear trend lines. The Pearson correlation between two scores increases when h increased.}
	\label{fig:scatterplot}
	\vspace{-10pt}
\end{figure}
%%%%%%%%%%%%%%%% end

%---figure vary number of hops with varying embedding sizes
\begin{figure*}[ht]
	\begin{subfigure}{0.162\textwidth}
		\centering
		\includegraphics[width=\textwidth]{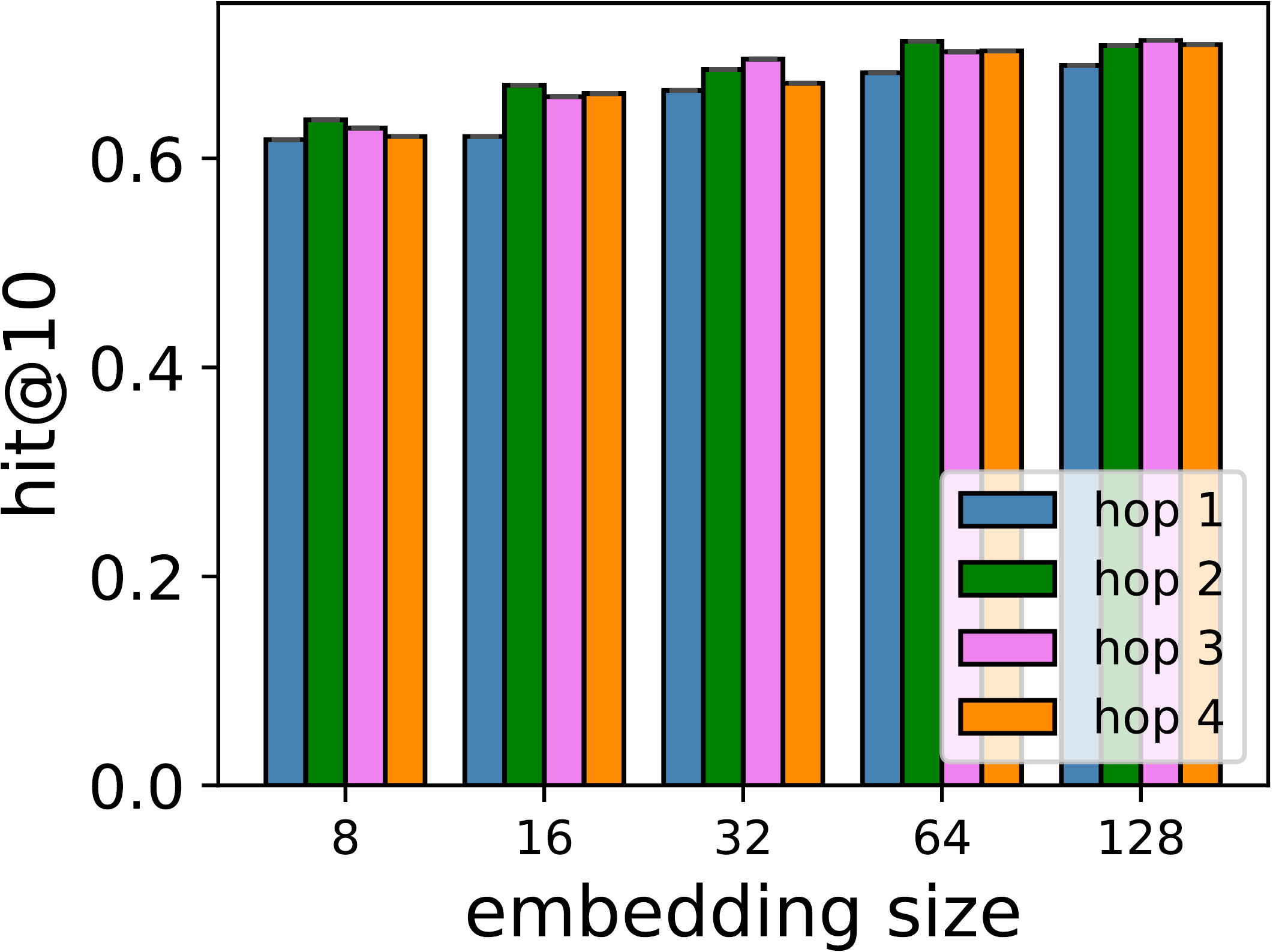}
		\caption{ML-100K.}
		\label{fig:ml100k-vary-hop-embed}
	\end{subfigure}
	\begin{subfigure}{0.162\textwidth}
		\centering
		\includegraphics[width=\textwidth]{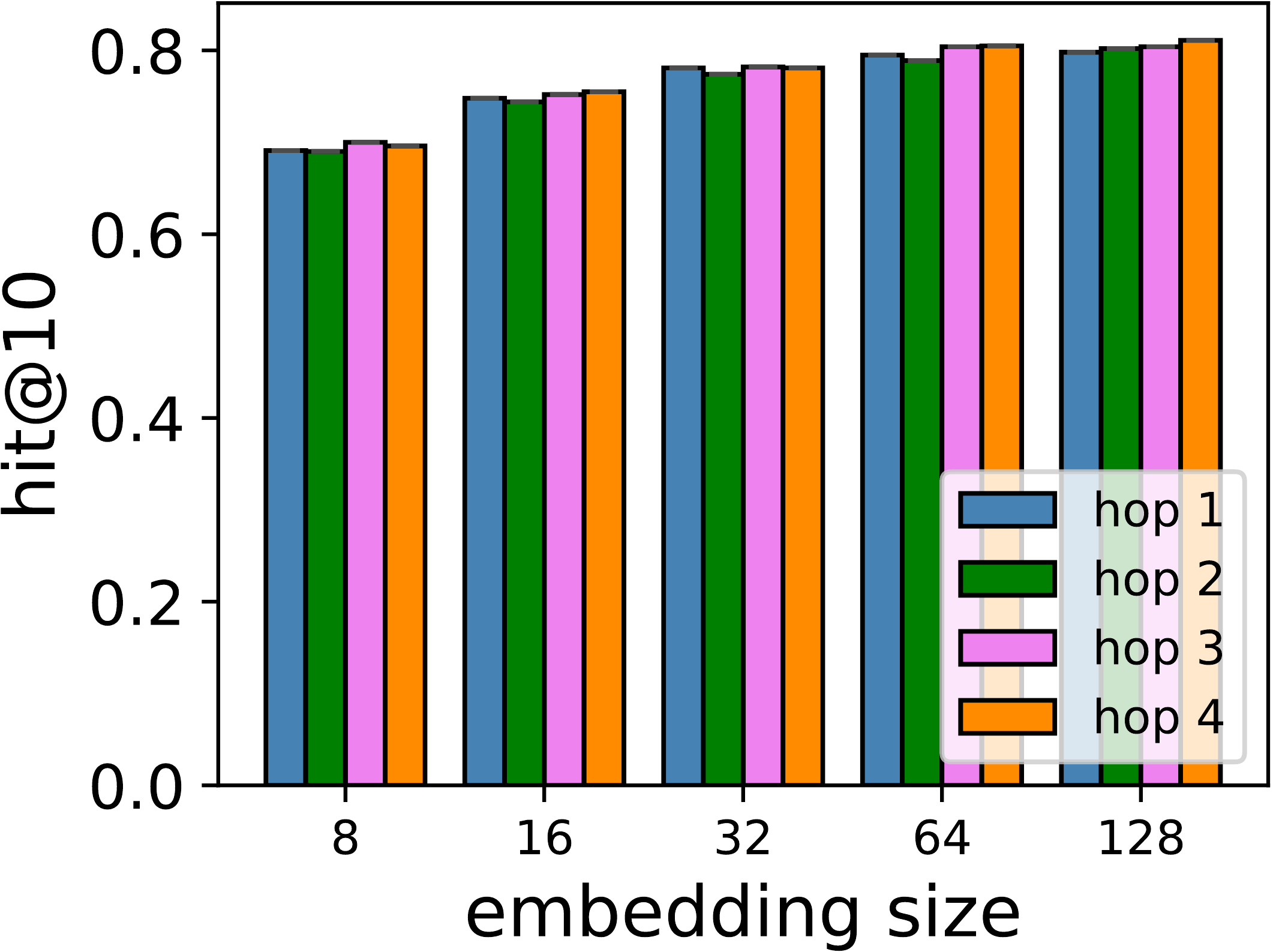}
		\caption{ML-1M.}
		\label{fig:ml1m-vary-hop-embed}
	\end{subfigure}
	\begin{subfigure}{0.162\textwidth}
		\centering
		\includegraphics[width=\textwidth]{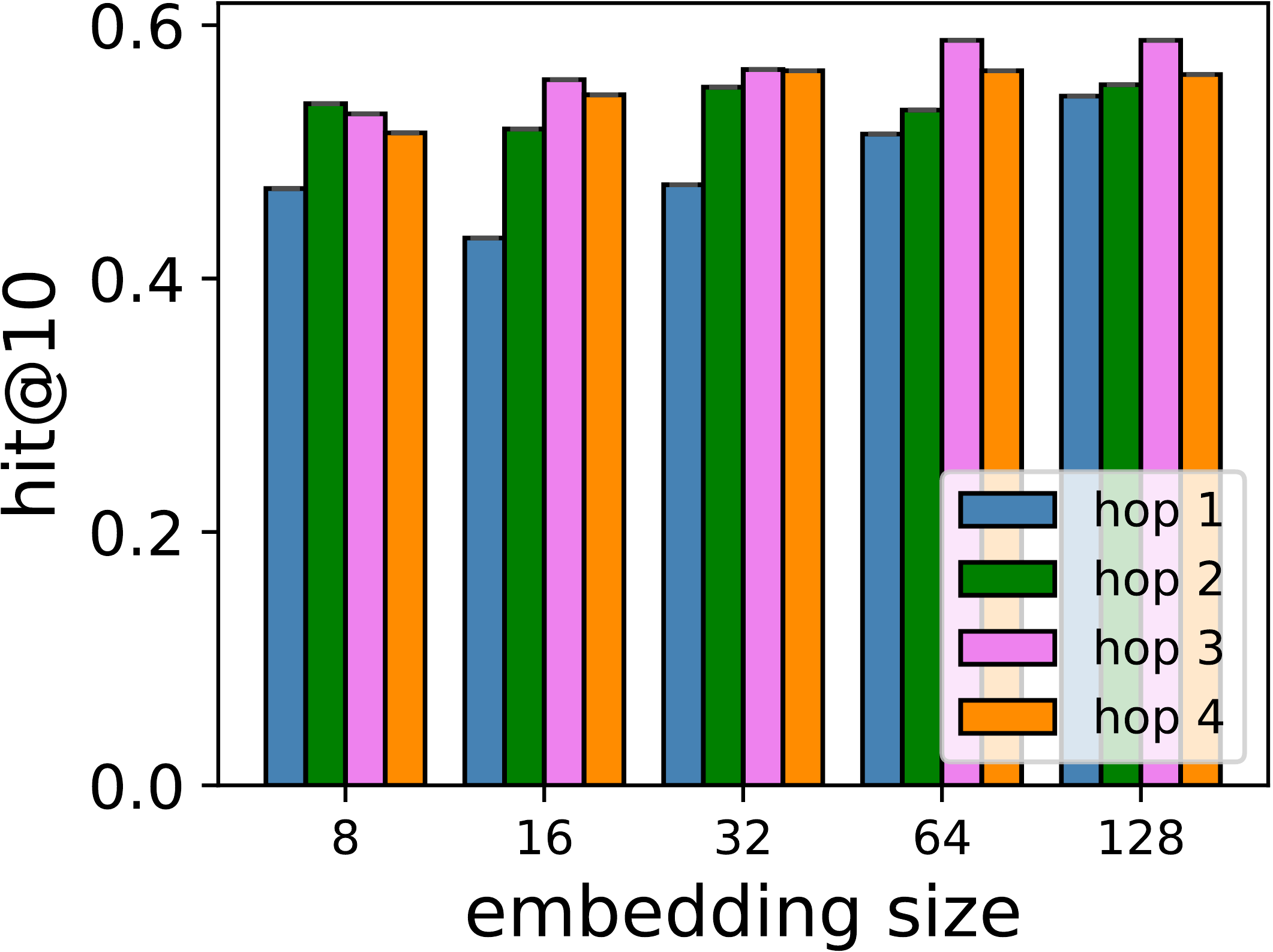}
		\caption{Netflix.}
		\label{fig:netflix-vary-hop-embed}
	\end{subfigure}
	\begin{subfigure}{0.162\textwidth}
		\centering
		\includegraphics[width=\textwidth]{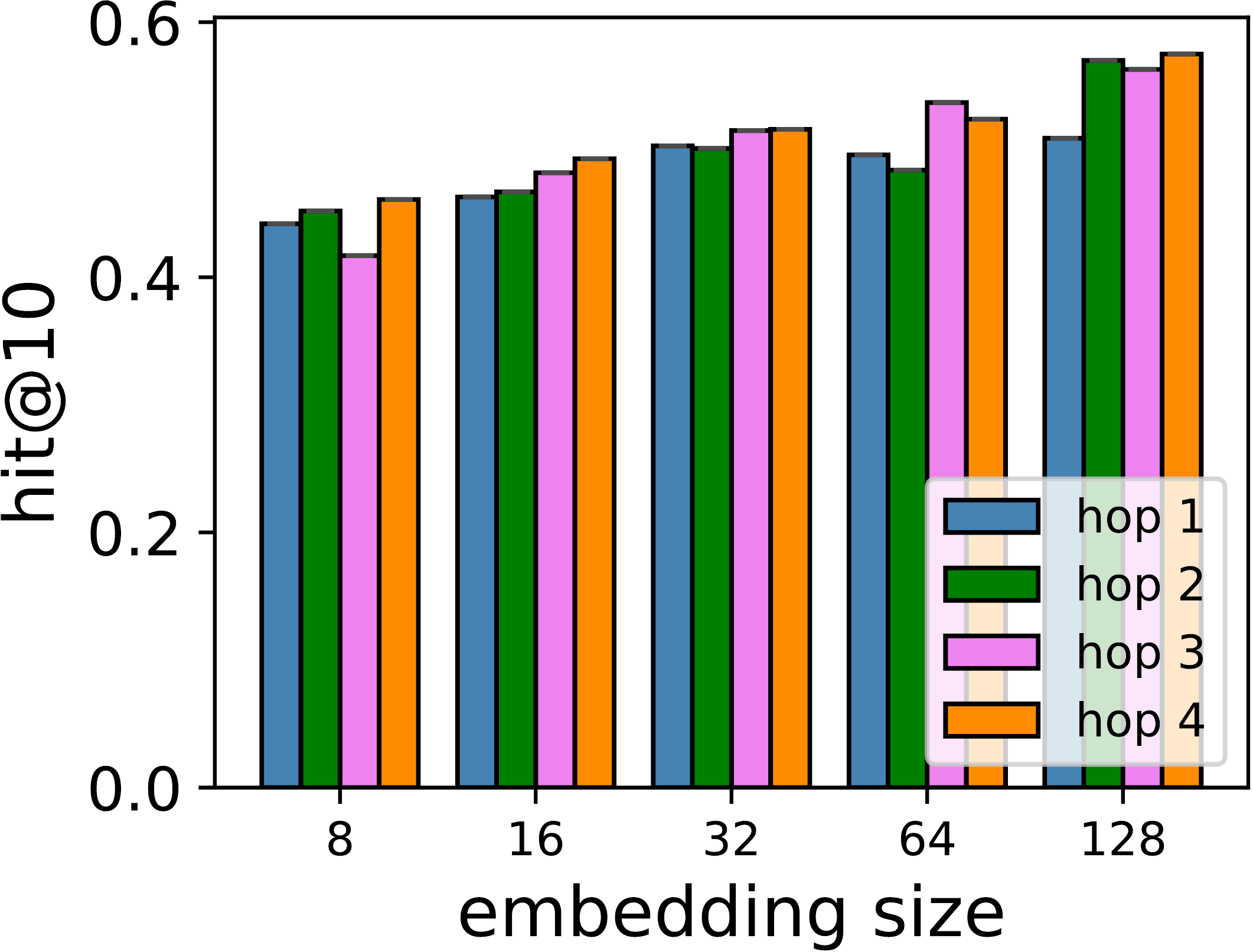}
		\caption{Epinions.}
		\label{fig:epinions-full-vary-hop-embed}
	\end{subfigure}
	\begin{subfigure}{0.162\textwidth}
		\centering
		\includegraphics[width=\textwidth]{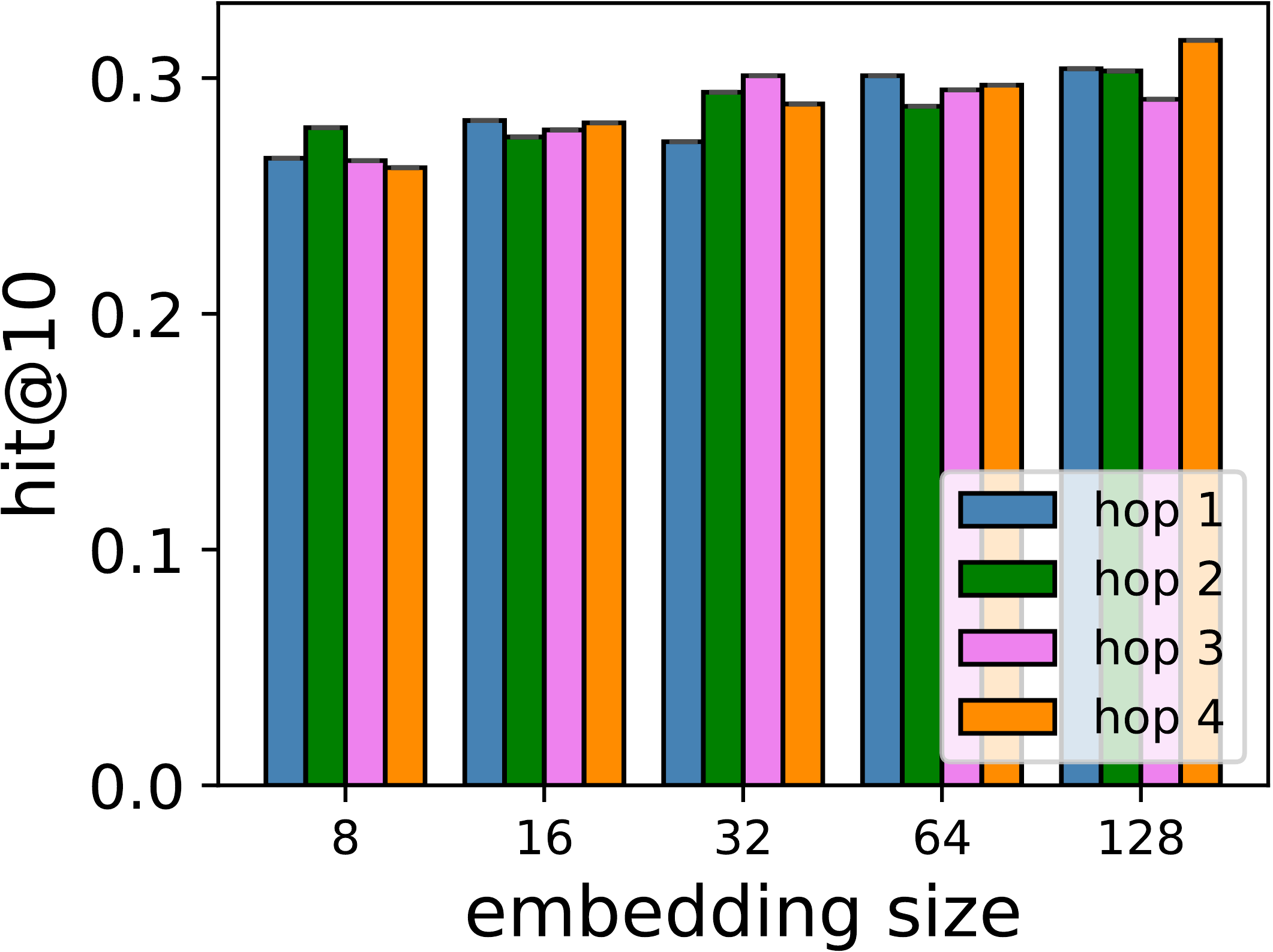}
		\caption{IJCAI-15.}
		\label{fig:ijcai-2015-20k-vary-hop-embed}
	\end{subfigure}
	\begin{subfigure}{0.162\textwidth}
		\centering
		\includegraphics[width=\textwidth]{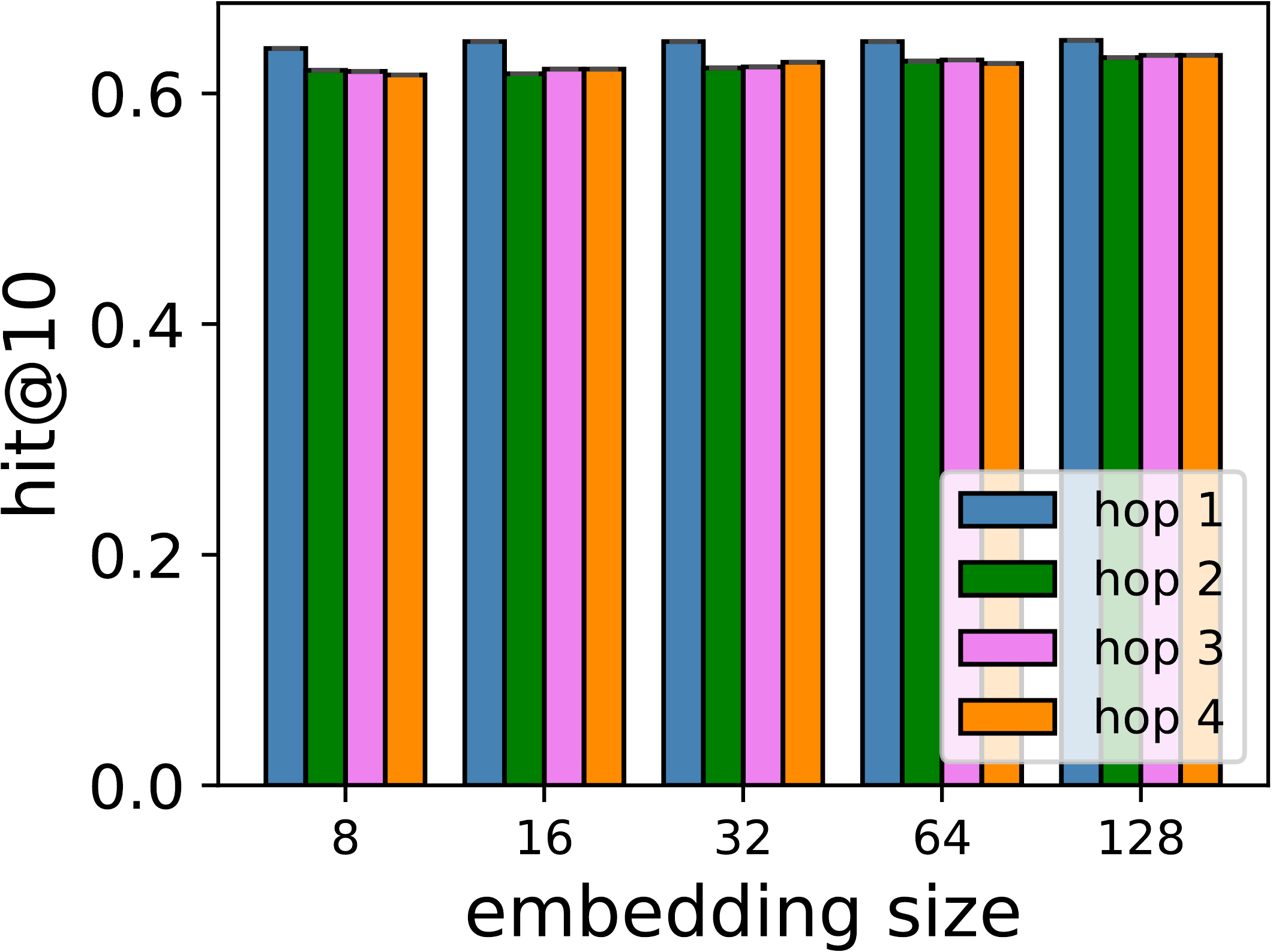}
		\caption{TaFeng.}
		\label{fig:tafeng-vary-hop-embed}
	\end{subfigure}
	\vspace{-5pt}
	\caption{Comparison of varying the number of hops regarding different embeddings sizes in the six datasets.}
	\label{fig:varying-hop-embedding}
	\vspace{-5pt}
\end{figure*}
%--

\vspace{-5pt}
\subsection{RQ2: Understanding our multi-hop personalized metric-based attention design?}
\begin{figure}[t]
	\centering
	\includegraphics[width=0.9\linewidth]{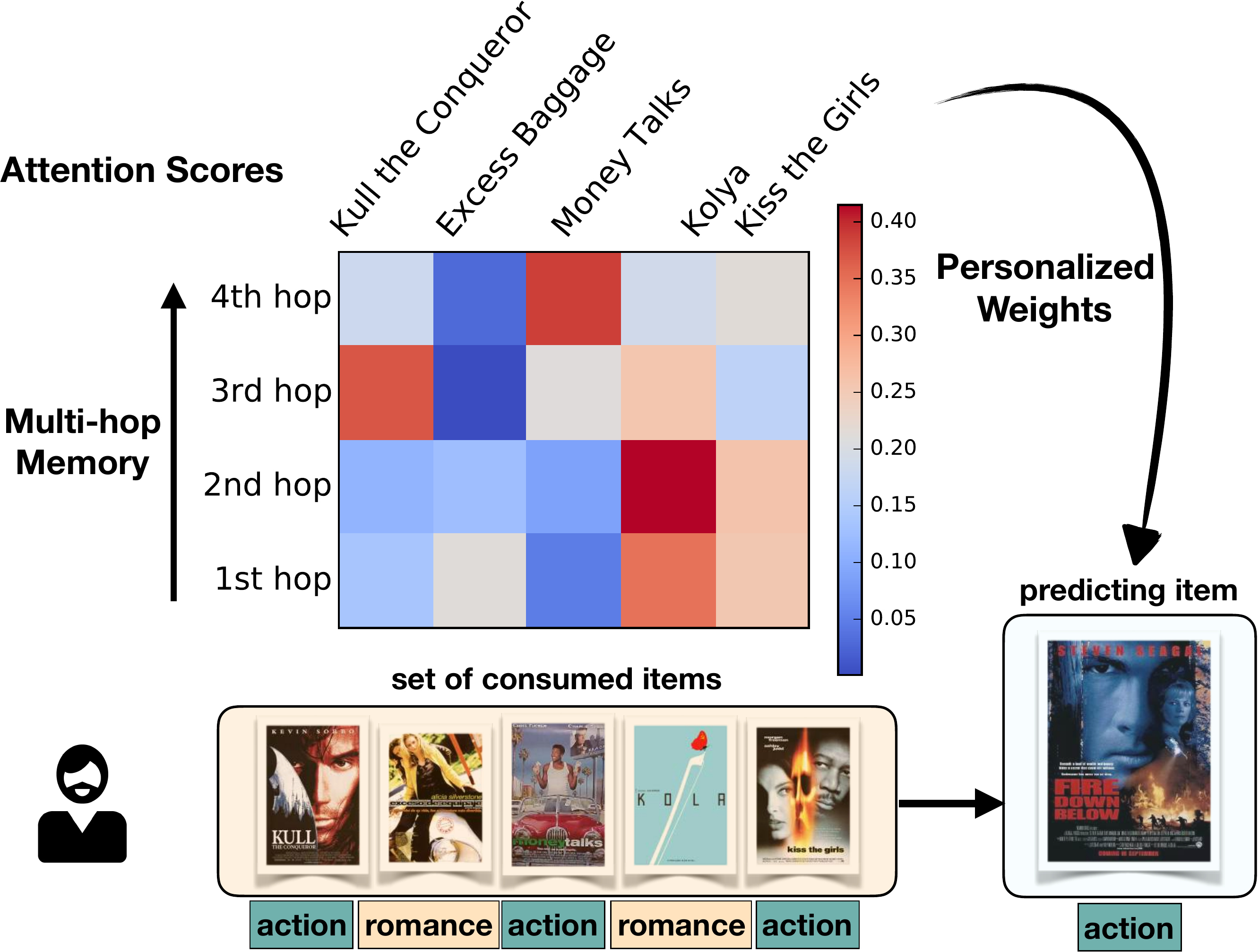}
	\vspace{-5pt}
	\caption{Multi-hop Attention visualization.}
	\label{fig:atteion-visualization}
	\vspace{-10pt}
\end{figure}

In the previous section, we see that our  models outperformed many strong baselines in \textbf{six} different datasets of the \textbf{two} different recommendation problems.
In this part, we explore why did we achieve those better results? As ``attention is all you need'' \cite{vaswani2017attention}, the core reason brought us an surpassed performance accredit to the  metric-based attention which are further refined via multi-hop design. Therefore, we want to explore quantitatively and qualitatively how our attention with multi-hop design worked by answering two smaller research questions: (i) what did our metric-based attention with multi-hop design learn?, (ii) did the metric-based attention with multi-hop design improve recommendation results? Without a special mention, since our SDMR model just learned a combination between SDP and SDM without re-learning the learned-already parameters in SDP and SDM, we explore SDM in this section to understand how attention with multi-hop design works. Note that we conduct this analysis for ML-100k only due to space limitation and the availability of movies genre in ML-100k (for visualization in Figure \ref{fig:atteion-visualization}).

\noindent\textbf{What did our metric-based attention with multi-hop design learn?} To answer this question, we first measure the \emph{point-wise mutual information} (PMI) between two certain items $j$ and $k$ as:
\vspace{-5pt}
\begin{equation}
\label{equa:pmi}
PMI(j, k) = log \frac{P(j,k)}{P(j) \times P(k)}
\end{equation}
where $P(j,k)$ is the joint probability between two items $j$ and $k$, which shows how likely $j$ and $k$ are co-preferred ($P(j,k)=\frac{\#(j,k)}{|D|}$, where $D$ denotes a collection of all item-item pairs, and $|D|$ is the total number of item-item co-occurrence pairs in D). Similarly, $P(j)$ and $P(k)$ are the probabilities of the item $j$ and $k$ in D, respectively (e.g. $P(j)=\frac{\#(j)}{|D|}$, $P(k)=\frac{\#(k)}{|D|}$). Intuitively, a PMI score between two items shows how likely the two items are co-purchased/co-preferred. The higher the PMI score between $j$ and $k$ is, the more likely the user will purchase $j$ if $k$ was purchased before.

We denote SDM-h is the SDM model with h hops. Now, given a target item $j$ and the user's context items $k$, \emph{SDM-h} will assign attentive scores for all ($j, k$) pairs. We also get PMI scores (from Eq.~(\ref{equa:pmi})) of ($j, k$) pairs%, which somewhat show how similar $j$ and $k$ are.
. Next, we plot a scatter plot of PMI scores and attentive scores for all ($j, k$) pairs to see the relationship between the two scores. Our results for ML-100k dataset is shown in Figure~\ref{fig:scatterplot}.

In Figure~\ref{fig:scatterplot}, the Pearson correlation between PMI scores and attentive scores are 0.059, 0.097, 0.143, and 0.146 for SDM-1, SDM-2, SDM-3 SDM-4, respectively. It indicates that as we increase the number of hops in SDM model, PMI scores and attentive scores are more positively correlated. In another word, as we increase the number of hops, our metric-based attention with multi-hop design will assign higher weights for co-purchased items, which is what we desire. 
Furthermore, scatter plots in Figure~\ref{fig:pmi-attention-hop1} presents that there is a high density of points with small attentive scores. This indicates that attention in SDM-1 is distributed to several items (which is somewhat close to equally focusing on context items). However, when we increase the number of hops $h$, the density spreads up to the top, indicating that the model tends to give a higher attention to some context items, which can be more relevant than others. This observation is consistent with ``learning to attend'' in \cite{xu2015show,bahdanau2014neural}.

%We also observe a similar phenomenon in other datasets.

%Next, we measure the Pearson correlation between the two coordinates of all points $p$ generated by SDM\_h. Our idea is the higher the Pearson correlation scores are, the higher the attention scores the model gives for similar items. We then show the Pearson correlation scores of SDM-h with $h \in$ \{1, 2, 3, 4\} in Table~\ref{table:pearsonCorrelation}

%\begin{table}[]
%    \centering
%	\tiny
%	\caption{Pearson correlation scores between PMI and attention scores between item-item pairs.}
%	%\vspace{-5pt}
%	\label{table:pearsonCorrelation}
%	\resizebox{0.45\textwidth}{!}{
%    \begin{tabular}{lcccc}
%    \toprule
%    \multirow{2}{*}{Dataset} & \multicolumn{4}{c}{Pearson Correlation} \\
%    \cmidrule{2-5}                         & SDM-1    & SDM-2     & SDM-3   & SDM-4   \\
%    \midrule
%    ml100k                   & 0.068    & 0.101     & 0.155   & 0.157   \\
%    ml1m                     & 0.081    & 0.151     & 0.165   & 0.121   \\
%    netflix                  & 0.023    & 0.033     & 0.074   & 0.019   \\
%    epinions                 & 0.047    & 0.062     & 0.050   & 0.054   \\
%    ijcai-15                 &          &           &         &         \\
%    tafeng                   & 0.052    & 0.031     & 0.023   & 0.019   \\
%    \bottomrule
%    \end{tabular}
%}
%\end{table}

\noindent\textbf{Did the metric-based attention with multi-hop design improve recommendation results?} We answer this research question by showing the results of SDM model when varying number of hops $h$ from \{1, 2, 3, 4\} with different embedding sizes and visualize attention scores of SDM-h with a random observation as follows:

\noindent\textbf{Varying number of hops with different embedding sizes:}  The performance of SDM-h regarding \emph{hit}@10 with $h$ from \{1, 2, 3, 4\} and embedding size from \{8, 16, 32, 64, 128\} is presented in Figure \ref{fig:varying-hop-embedding}. We see that more hops tend to give additional improvement in all 6 datasets, except in Tafeng dataset where SDM with more hops over-fitted. In ML-100k and ML-1M, the optimal number of hops are 3 or 4. In Netflix, SDM with 3 hops performed well. In Epinions and IJCAI-15, SDM-4 tends to achieve better results.
Overall, the selection of the number of hops depends on the dataset complexity, and it varies from datasets to datasets.

%% -- end
\noindent\textbf{Attention Visualization:}
Lastly, to visualize how the personalized metric-based attention with multi-hop design works, we chose one user from ML-100K data. The learned weights at each hop of SDM is shown in Figure~\ref{fig:atteion-visualization}. The target item in this example is an action movie called \emph{Fire Down Below} (1997).
%We randomly select one user and show the personalized attentive scores from each of her consumed items \textit{w.r.t.} the predicting item.
%The user's 5 recent watched movies in order are: \emph{Kull the Conqueror (1997)} - an action movie,  \emph{Excess Baggage (1997)} - a drama/romance movie, \emph{Money Talks (1997)} - an action movie, \emph{Kolya (1996)} - a drama movie, \emph{Kiss the Girls (1997)} - an action movie.
The first two hops of SDM assigned high weights to two romance movies, and the lowest score to the action movie \emph{Money Talks} (1997).
The 3rd-hop and 4th-hop attention refined the weights of movies to better reflect the correlations and similarities \textit{w.r.t} the target movie.
At last, \emph{Money Talks} (1997) was assigned with the highest weight $0.386$, and the total weights of two romance movies decreased to less than $0.2$. This result shows the effectiveness of our multi-hop SDM model.

%by increasing weights for action movies  decreasing weights for romance movies (especially \emph{Excess Baggage} (1997) with only $0.01$ at 4-th hop), which are less similar to the predicting item.
%At for \emph{Money Talks} (1997), which has a same genre with the predicting item.

%The predicting item is \emph{Raise the Red Lantern (1991)} - a drama/romance movie. The five recent movies the user watched in order are: \emph{E.T. the Extra-Terrestrial (1982)} - a sci-fi movie, \emph{My Fair Lady (1964)} - a drama movie, \emph{Farewell My Concubine (1993)} - a drama movie, \emph{Three Colors: Red (1994)} - a drama/romance movie, and \emph{The Shawshank Redemption (1994)} - a drama movie.
%We observe that our personalized attention module produces higher scores for drama movies, which have a same genre with the predicting movie, while assigned a very small weight for \emph{E.T. the Extra-Terrestrial (1982)}, which is an unrelated movie. Furthermore, the personalized attention scores are spread out, showing various contributions of different items toward the predicting item.

\section{Conclusion}
\label{sec:conclusion}
In this paper, we have studied the top-$k$ recommendation problem in a signed distance learning perspective.
Different from previous works, we have considered two independent signed distance models for measuring user-item and item-item similarities respectively via deep neural networks.
%The proposed SDP learns a non-linear metric in a user-item latent space. while
%SDM learns a personalized item-item distance with soft attention.
%Then the two networks are combined for a compounded distance metric approximator called SDMR.
Extensive experiments have been performed on six real-world datasets in general recommendation and shopping basket-based recommendation task. We presented that our proposed SDMR outperformed \textbf{ten} baselines in all two recommendation tasks. To an extension, future works can integrate position embeddings \cite{vaswani2017attention} in our models, which give our models a sense of which position they are dealing with, which can further improve our model's performance when rigid orders of items are available.

\section*{Acknowledgment}
This work was supported in part by NSF grant CNS-1755536, Google Faculty Research Award, Microsoft Azure Research Award, AWS Cloud Credits for Research, and Google Cloud. Any opinions, findings and conclusions or recommendations expressed in this material are the author(s) and do not necessarily reflect those of the sponsors.

\bibliographystyle{ACM-Reference-Format}
\balance 
\bibliography{ref}


\begin{thebibliography}{63}

%%% the Web address.

\ifx \showCODEN    \undefined \def \showCODEN     #1{\unskip}     \fi
\ifx \showDOI      \undefined \def \showDOI       #1{#1}\fi
\ifx \showISBNx    \undefined \def \showISBNx     #1{\unskip}     \fi
\ifx \showISBNxiii \undefined \def \showISBNxiii  #1{\unskip}     \fi
\ifx \showISSN     \undefined \def \showISSN      #1{\unskip}     \fi
\ifx \showLCCN     \undefined \def \showLCCN      #1{\unskip}     \fi
\ifx \shownote     \undefined \def \shownote      #1{#1}          \fi
\ifx \showarticletitle \undefined \def \showarticletitle #1{#1}   \fi
\ifx \showURL      \undefined \def \showURL       {\relax}        \fi
% invisible to TeX
\providecommand\bibfield[2]{#2}
\providecommand\bibinfo[2]{#2}
\providecommand\natexlab[1]{#1}
\providecommand\showeprint[2][]{arXiv:#2}

\bibitem[\protect\citeauthoryear{Aggarwal}{Aggarwal}{2016}]%
        {aggarwal2016recommender}
\bibfield{author}{\bibinfo{person}{Charu~C Aggarwal}.}
  \bibinfo{year}{2016}\natexlab{}.
\newblock \bibinfo{booktitle}{\emph{Recommender systems}}.
\newblock \bibinfo{publisher}{Springer}.
\newblock


\bibitem[\protect\citeauthoryear{Bahdanau, Cho, and Bengio}{Bahdanau
  et~al\mbox{.}}{2014}]%
        {bahdanau2014neural}
\bibfield{author}{\bibinfo{person}{Dzmitry Bahdanau},
  \bibinfo{person}{Kyunghyun Cho}, {and} \bibinfo{person}{Yoshua Bengio}.}
  \bibinfo{year}{2014}\natexlab{}.
\newblock \showarticletitle{Neural machine translation by jointly learning to
  align and translate}.
\newblock \bibinfo{journal}{\emph{arXiv preprint arXiv:1409.0473}}
  (\bibinfo{year}{2014}).
\newblock


\bibitem[\protect\citeauthoryear{Billsus and Pazzani}{Billsus and
  Pazzani}{2000}]%
        {billsus2000user}
\bibfield{author}{\bibinfo{person}{Daniel Billsus} {and}
  \bibinfo{person}{Michael~J Pazzani}.} \bibinfo{year}{2000}\natexlab{}.
\newblock \showarticletitle{User modeling for adaptive news access}.
\newblock \bibinfo{journal}{\emph{User modeling and user-adapted interaction}}
  \bibinfo{volume}{10}, \bibinfo{number}{2-3} (\bibinfo{year}{2000}),
  \bibinfo{pages}{147--180}.
\newblock


\bibitem[\protect\citeauthoryear{Chen, Zhang, He, Nie, Liu, and Chua}{Chen
  et~al\mbox{.}}{2017}]%
        {chen2017attentive}
\bibfield{author}{\bibinfo{person}{Jingyuan Chen}, \bibinfo{person}{Hanwang
  Zhang}, \bibinfo{person}{Xiangnan He}, \bibinfo{person}{Liqiang Nie},
  \bibinfo{person}{Wei Liu}, {and} \bibinfo{person}{Tat-Seng Chua}.}
  \bibinfo{year}{2017}\natexlab{}.
\newblock \showarticletitle{Attentive collaborative filtering: Multimedia
  recommendation with item-and component-level attention}. In
  \bibinfo{booktitle}{\emph{Proceedings of the 40th International ACM SIGIR
  conference on Research and Development in Information Retrieval}}.
  \bibinfo{pages}{335--344}.
\newblock


\bibitem[\protect\citeauthoryear{Choi, Cho, and Bengio}{Choi
  et~al\mbox{.}}{2018}]%
        {choi2018fine}
\bibfield{author}{\bibinfo{person}{Heeyoul Choi}, \bibinfo{person}{Kyunghyun
  Cho}, {and} \bibinfo{person}{Yoshua Bengio}.}
  \bibinfo{year}{2018}\natexlab{}.
\newblock \showarticletitle{Fine-grained attention mechanism for neural machine
  translation}.
\newblock \bibinfo{journal}{\emph{Neurocomputing}}  \bibinfo{volume}{284}
  (\bibinfo{year}{2018}), \bibinfo{pages}{171--176}.
\newblock


\bibitem[\protect\citeauthoryear{Deshpande and Karypis}{Deshpande and
  Karypis}{2004}]%
        {deshpande2004item}
\bibfield{author}{\bibinfo{person}{Mukund Deshpande} {and}
  \bibinfo{person}{George Karypis}.} \bibinfo{year}{2004}\natexlab{}.
\newblock \showarticletitle{Item-based top-n recommendation algorithms}.
\newblock \bibinfo{journal}{\emph{ACM Transactions on Information Systems}}
  \bibinfo{volume}{22}, \bibinfo{number}{1} (\bibinfo{year}{2004}),
  \bibinfo{pages}{143--177}.
\newblock


\bibitem[\protect\citeauthoryear{Ebesu, Shen, and Fang}{Ebesu
  et~al\mbox{.}}{2018}]%
        {ebesu2018collaborative}
\bibfield{author}{\bibinfo{person}{Travis Ebesu}, \bibinfo{person}{Bin Shen},
  {and} \bibinfo{person}{Yi Fang}.} \bibinfo{year}{2018}\natexlab{}.
\newblock \showarticletitle{Collaborative Memory Network for Recommendation
  Systems}. In \bibinfo{booktitle}{\emph{Proceedings of the 41st ACM
  International Conference on Research and Development in Information
  Retrieval}}.
\newblock


\bibitem[\protect\citeauthoryear{Feng, Li, Zeng, Cong, Chee, and Yuan}{Feng
  et~al\mbox{.}}{2015}]%
        {feng2015personalized}
\bibfield{author}{\bibinfo{person}{Shanshan Feng}, \bibinfo{person}{Xutao Li},
  \bibinfo{person}{Yifeng Zeng}, \bibinfo{person}{Gao Cong},
  \bibinfo{person}{Yeow~Meng Chee}, {and} \bibinfo{person}{Quan Yuan}.}
  \bibinfo{year}{2015}\natexlab{}.
\newblock \showarticletitle{Personalized Ranking Metric Embedding for Next New
  POI Recommendation}. In \bibinfo{booktitle}{\emph{International Joint
  Conference on Artificial Intelligence}}, Vol.~\bibinfo{volume}{15}.
  \bibinfo{pages}{2069--2075}.
\newblock


\bibitem[\protect\citeauthoryear{He, Kang, and McAuley}{He
  et~al\mbox{.}}{2017a}]%
        {he2017translation}
\bibfield{author}{\bibinfo{person}{Ruining He}, \bibinfo{person}{Wang-Cheng
  Kang}, {and} \bibinfo{person}{Julian McAuley}.}
  \bibinfo{year}{2017}\natexlab{a}.
\newblock \showarticletitle{Translation-based recommendation}. In
  \bibinfo{booktitle}{\emph{Proceedings of the Eleventh ACM Conference on
  Recommender Systems}}. \bibinfo{pages}{161--169}.
\newblock


\bibitem[\protect\citeauthoryear{He and McAuley}{He and McAuley}{2016a}]%
        {he2016fusing}
\bibfield{author}{\bibinfo{person}{Ruining He} {and} \bibinfo{person}{Julian
  McAuley}.} \bibinfo{year}{2016}\natexlab{a}.
\newblock \showarticletitle{Fusing similarity models with markov chains for
  sparse sequential recommendation}. In \bibinfo{booktitle}{\emph{Data Mining
  (ICDM), 2016 IEEE 16th International Conference on}}.
  \bibinfo{pages}{191--200}.
\newblock


\bibitem[\protect\citeauthoryear{He and McAuley}{He and McAuley}{2016b}]%
        {he2016ups}
\bibfield{author}{\bibinfo{person}{Ruining He} {and} \bibinfo{person}{Julian
  McAuley}.} \bibinfo{year}{2016}\natexlab{b}.
\newblock \showarticletitle{Ups and downs: Modeling the visual evolution of
  fashion trends with one-class collaborative filtering}. In
  \bibinfo{booktitle}{\emph{proceedings of the 25th international conference on
  world wide web}}. \bibinfo{pages}{507--517}.
\newblock


\bibitem[\protect\citeauthoryear{He, Liao, Zhang, Nie, Hu, and Chua}{He
  et~al\mbox{.}}{2017b}]%
        {he2017neural}
\bibfield{author}{\bibinfo{person}{Xiangnan He}, \bibinfo{person}{Lizi Liao},
  \bibinfo{person}{Hanwang Zhang}, \bibinfo{person}{Liqiang Nie},
  \bibinfo{person}{Xia Hu}, {and} \bibinfo{person}{Tat-Seng Chua}.}
  \bibinfo{year}{2017}\natexlab{b}.
\newblock \showarticletitle{Neural collaborative filtering}. In
  \bibinfo{booktitle}{\emph{Proceedings of the 26th International Conference on
  World Wide Web}}. \bibinfo{pages}{173--182}.
\newblock


\bibitem[\protect\citeauthoryear{He, Zhang, Kan, and Chua}{He
  et~al\mbox{.}}{2016}]%
        {he2016fast}
\bibfield{author}{\bibinfo{person}{Xiangnan He}, \bibinfo{person}{Hanwang
  Zhang}, \bibinfo{person}{Min-Yen Kan}, {and} \bibinfo{person}{Tat-Seng
  Chua}.} \bibinfo{year}{2016}\natexlab{}.
\newblock \showarticletitle{Fast matrix factorization for online recommendation
  with implicit feedback}. In \bibinfo{booktitle}{\emph{Proceedings of the 39th
  International ACM SIGIR conference on Research and Development in Information
  Retrieval}}. \bibinfo{pages}{549--558}.
\newblock


\bibitem[\protect\citeauthoryear{Hidasi, Karatzoglou, Baltrunas, and
  Tikk}{Hidasi et~al\mbox{.}}{2015}]%
        {hidasi2015session}
\bibfield{author}{\bibinfo{person}{Bal{\'a}zs Hidasi},
  \bibinfo{person}{Alexandros Karatzoglou}, \bibinfo{person}{Linas Baltrunas},
  {and} \bibinfo{person}{Domonkos Tikk}.} \bibinfo{year}{2015}\natexlab{}.
\newblock \showarticletitle{Session-based recommendations with recurrent neural
  networks}.
\newblock \bibinfo{journal}{\emph{arXiv preprint arXiv:1511.06939}}
  (\bibinfo{year}{2015}).
\newblock


\bibitem[\protect\citeauthoryear{Hsieh, Yang, Cui, Lin, Belongie, and
  Estrin}{Hsieh et~al\mbox{.}}{2017}]%
        {hsieh2017collaborative}
\bibfield{author}{\bibinfo{person}{Cheng-Kang Hsieh}, \bibinfo{person}{Longqi
  Yang}, \bibinfo{person}{Yin Cui}, \bibinfo{person}{Tsung-Yi Lin},
  \bibinfo{person}{Serge Belongie}, {and} \bibinfo{person}{Deborah Estrin}.}
  \bibinfo{year}{2017}\natexlab{}.
\newblock \showarticletitle{Collaborative metric learning}. In
  \bibinfo{booktitle}{\emph{Proceedings of the 26th International Conference on
  World Wide Web}}. \bibinfo{pages}{193--201}.
\newblock


\bibitem[\protect\citeauthoryear{Hu, Koren, and Volinsky}{Hu
  et~al\mbox{.}}{2008}]%
        {hu2008collaborative}
\bibfield{author}{\bibinfo{person}{Yifan Hu}, \bibinfo{person}{Yehuda Koren},
  {and} \bibinfo{person}{Chris Volinsky}.} \bibinfo{year}{2008}\natexlab{}.
\newblock \showarticletitle{Collaborative filtering for implicit feedback
  datasets}. In \bibinfo{booktitle}{\emph{Data Mining, 2008. ICDM'08. Eighth
  IEEE International Conference on}}. \bibinfo{pages}{263--272}.
\newblock


\bibitem[\protect\citeauthoryear{Kim, Park, Oh, Lee, and Yu}{Kim
  et~al\mbox{.}}{2016}]%
        {kim2016convolutional}
\bibfield{author}{\bibinfo{person}{Donghyun Kim}, \bibinfo{person}{Chanyoung
  Park}, \bibinfo{person}{Jinoh Oh}, \bibinfo{person}{Sungyoung Lee}, {and}
  \bibinfo{person}{Hwanjo Yu}.} \bibinfo{year}{2016}\natexlab{}.
\newblock \showarticletitle{Convolutional matrix factorization for document
  context-aware recommendation}. In \bibinfo{booktitle}{\emph{Proceedings of
  the 10th ACM Conference on Recommender Systems}}. \bibinfo{pages}{233--240}.
\newblock


\bibitem[\protect\citeauthoryear{Kingma and Ba}{Kingma and Ba}{2014}]%
        {kingma2014adam}
\bibfield{author}{\bibinfo{person}{Diederik~P Kingma} {and}
  \bibinfo{person}{Jimmy Ba}.} \bibinfo{year}{2014}\natexlab{}.
\newblock \showarticletitle{Adam: A method for stochastic optimization}.
\newblock \bibinfo{journal}{\emph{arXiv preprint arXiv:1412.6980}}
  (\bibinfo{year}{2014}).
\newblock


\bibitem[\protect\citeauthoryear{Koren}{Koren}{2008}]%
        {koren2008factorization}
\bibfield{author}{\bibinfo{person}{Yehuda Koren}.}
  \bibinfo{year}{2008}\natexlab{}.
\newblock \showarticletitle{Factorization meets the neighborhood: a
  multifaceted collaborative filtering model}. In
  \bibinfo{booktitle}{\emph{Proceedings of the 14th ACM SIGKDD international
  conference on Knowledge discovery and data mining}}.
  \bibinfo{pages}{426--434}.
\newblock


\bibitem[\protect\citeauthoryear{Koren}{Koren}{2009}]%
        {koren2009collaborative}
\bibfield{author}{\bibinfo{person}{Yehuda Koren}.}
  \bibinfo{year}{2009}\natexlab{}.
\newblock \showarticletitle{Collaborative filtering with temporal dynamics}. In
  \bibinfo{booktitle}{\emph{Proceedings of the 15th ACM SIGKDD international
  conference on Knowledge discovery and data mining}}.
  \bibinfo{pages}{447--456}.
\newblock


\bibitem[\protect\citeauthoryear{Koren}{Koren}{2010}]%
        {koren2010collaborative}
\bibfield{author}{\bibinfo{person}{Yehuda Koren}.}
  \bibinfo{year}{2010}\natexlab{}.
\newblock \showarticletitle{Collaborative filtering with temporal dynamics}.
\newblock \bibinfo{journal}{\emph{Commun. ACM}} \bibinfo{volume}{53},
  \bibinfo{number}{4} (\bibinfo{year}{2010}), \bibinfo{pages}{89--97}.
\newblock


\bibitem[\protect\citeauthoryear{Lang}{Lang}{1995}]%
        {lang1995newsweeder}
\bibfield{author}{\bibinfo{person}{Ken Lang}.} \bibinfo{year}{1995}\natexlab{}.
\newblock \showarticletitle{Newsweeder: Learning to filter netnews}.
\newblock In \bibinfo{booktitle}{\emph{International Conference on Machine
  Learning}}. \bibinfo{pages}{331--339}.
\newblock


\bibitem[\protect\citeauthoryear{Lawrence and Urtasun}{Lawrence and
  Urtasun}{2009}]%
        {lawrence2009non}
\bibfield{author}{\bibinfo{person}{Neil~D Lawrence} {and}
  \bibinfo{person}{Raquel Urtasun}.} \bibinfo{year}{2009}\natexlab{}.
\newblock \showarticletitle{Non-linear matrix factorization with Gaussian
  processes}. In \bibinfo{booktitle}{\emph{Proceedings of the 26th Annual
  International Conference on Machine Learning}}. \bibinfo{pages}{601--608}.
\newblock


\bibitem[\protect\citeauthoryear{Li, Kawale, and Fu}{Li et~al\mbox{.}}{2015}]%
        {li2015deep}
\bibfield{author}{\bibinfo{person}{Sheng Li}, \bibinfo{person}{Jaya Kawale},
  {and} \bibinfo{person}{Yun Fu}.} \bibinfo{year}{2015}\natexlab{}.
\newblock \showarticletitle{Deep collaborative filtering via marginalized
  denoising auto-encoder}. In \bibinfo{booktitle}{\emph{Proceedings of the 24th
  ACM International on Conference on Information and Knowledge Management}}.
  \bibinfo{pages}{811--820}.
\newblock


\bibitem[\protect\citeauthoryear{Liang, Krishnan, Hoffman, and Jebara}{Liang
  et~al\mbox{.}}{2018}]%
        {liang2018variational}
\bibfield{author}{\bibinfo{person}{Dawen Liang}, \bibinfo{person}{Rahul~G
  Krishnan}, \bibinfo{person}{Matthew~D Hoffman}, {and} \bibinfo{person}{Tony
  Jebara}.} \bibinfo{year}{2018}\natexlab{}.
\newblock \showarticletitle{Variational Autoencoders for Collaborative
  Filtering}.
\newblock  (\bibinfo{year}{2018}).
\newblock


\bibitem[\protect\citeauthoryear{Lin, Feng, Santos, Yu, Xiang, Zhou, and
  Bengio}{Lin et~al\mbox{.}}{2017}]%
        {lin2017structured}
\bibfield{author}{\bibinfo{person}{Zhouhan Lin}, \bibinfo{person}{Minwei Feng},
  \bibinfo{person}{Cicero Nogueira~dos Santos}, \bibinfo{person}{Mo Yu},
  \bibinfo{person}{Bing Xiang}, \bibinfo{person}{Bowen Zhou}, {and}
  \bibinfo{person}{Yoshua Bengio}.} \bibinfo{year}{2017}\natexlab{}.
\newblock \showarticletitle{A structured self-attentive sentence embedding}.
\newblock \bibinfo{journal}{\emph{arXiv preprint arXiv:1703.03130}}
  (\bibinfo{year}{2017}).
\newblock


\bibitem[\protect\citeauthoryear{Linden, Smith, and York}{Linden
  et~al\mbox{.}}{2003}]%
        {linden2003amazon}
\bibfield{author}{\bibinfo{person}{Greg Linden}, \bibinfo{person}{Brent Smith},
  {and} \bibinfo{person}{Jeremy York}.} \bibinfo{year}{2003}\natexlab{}.
\newblock \showarticletitle{Amazon. com recommendations: Item-to-item
  collaborative filtering}.
\newblock \bibinfo{journal}{\emph{IEEE Internet computing}} \bibinfo{number}{1}
  (\bibinfo{year}{2003}), \bibinfo{pages}{76--80}.
\newblock


\bibitem[\protect\citeauthoryear{Liu and Perez}{Liu and Perez}{2017}]%
        {liu2017gated}
\bibfield{author}{\bibinfo{person}{Fei Liu} {and} \bibinfo{person}{Julien
  Perez}.} \bibinfo{year}{2017}\natexlab{}.
\newblock \showarticletitle{Gated end-to-end memory networks}. In
  \bibinfo{booktitle}{\emph{Proceedings of the 15th Conference of the European
  Chapter of the Association for Computational Linguistics: Volume 1, Long
  Papers}}, Vol.~\bibinfo{volume}{1}. \bibinfo{pages}{1--10}.
\newblock


\bibitem[\protect\citeauthoryear{Liu, Wu, and Wang}{Liu et~al\mbox{.}}{2017}]%
        {liu2017deepstyle}
\bibfield{author}{\bibinfo{person}{Qiang Liu}, \bibinfo{person}{Shu Wu}, {and}
  \bibinfo{person}{Liang Wang}.} \bibinfo{year}{2017}\natexlab{}.
\newblock \showarticletitle{DeepStyle: Learning user preferences for visual
  recommendation}. In \bibinfo{booktitle}{\emph{Proceedings of the 40th
  International ACM SIGIR Conference on Research and Development in Information
  Retrieval}}. \bibinfo{pages}{841--844}.
\newblock


\bibitem[\protect\citeauthoryear{Liu, Aggarwal, Li, Kong, Sun, and Sathe}{Liu
  et~al\mbox{.}}{2016}]%
        {liu2016kernelized}
\bibfield{author}{\bibinfo{person}{Xinyue Liu}, \bibinfo{person}{Chara
  Aggarwal}, \bibinfo{person}{Yu-Feng Li}, \bibinfo{person}{Xiaugnan Kong},
  \bibinfo{person}{Xinyuan Sun}, {and} \bibinfo{person}{Saket Sathe}.}
  \bibinfo{year}{2016}\natexlab{}.
\newblock \showarticletitle{Kernelized matrix factorization for collaborative
  filtering}. In \bibinfo{booktitle}{\emph{Proceedings of the 2016 SIAM
  International Conference on Data Mining}}. \bibinfo{pages}{378--386}.
\newblock


\bibitem[\protect\citeauthoryear{Lu, Dong, and Smyth}{Lu
  et~al\mbox{.}}{2018a}]%
        {lu2018coevolutionary}
\bibfield{author}{\bibinfo{person}{Yichao Lu}, \bibinfo{person}{Ruihai Dong},
  {and} \bibinfo{person}{Barry Smyth}.} \bibinfo{year}{2018}\natexlab{a}.
\newblock \showarticletitle{Coevolutionary Recommendation Model: Mutual
  Learning between Ratings and Reviews}. In
  \bibinfo{booktitle}{\emph{Proceedings of the 2018 World Wide Web Conference
  on World Wide Web}}. \bibinfo{pages}{773--782}.
\newblock


\bibitem[\protect\citeauthoryear{Lu, Dong, and Smyth}{Lu
  et~al\mbox{.}}{2018b}]%
        {lu2018convolutional}
\bibfield{author}{\bibinfo{person}{Yichao Lu}, \bibinfo{person}{Ruihai Dong},
  {and} \bibinfo{person}{Barry Smyth}.} \bibinfo{year}{2018}\natexlab{b}.
\newblock \showarticletitle{Convolutional Matrix Factorization for
  Recommendation Explanation}. In \bibinfo{booktitle}{\emph{Proceedings of the
  23rd International Conference on Intelligent User Interfaces Companion}}.
  \bibinfo{pages}{34}.
\newblock


\bibitem[\protect\citeauthoryear{Luong, Pham, and Manning}{Luong
  et~al\mbox{.}}{2015}]%
        {luong2015effective}
\bibfield{author}{\bibinfo{person}{Minh-Thang Luong}, \bibinfo{person}{Hieu
  Pham}, {and} \bibinfo{person}{Christopher~D Manning}.}
  \bibinfo{year}{2015}\natexlab{}.
\newblock \showarticletitle{Effective approaches to attention-based neural
  machine translation}.
\newblock \bibinfo{journal}{\emph{arXiv preprint arXiv:1508.04025}}
  (\bibinfo{year}{2015}).
\newblock


\bibitem[\protect\citeauthoryear{Ma, Kang, Wu, Wang, and Liu}{Ma
  et~al\mbox{.}}{2019}]%
        {ma2019gated}
\bibfield{author}{\bibinfo{person}{Chen Ma}, \bibinfo{person}{Peng Kang},
  \bibinfo{person}{Bin Wu}, \bibinfo{person}{Qinglong Wang}, {and}
  \bibinfo{person}{Xue Liu}.} \bibinfo{year}{2019}\natexlab{}.
\newblock \showarticletitle{Gated Attentive-Autoencoder for Content-Aware
  Recommendation}. In \bibinfo{booktitle}{\emph{Proceedings of the Twelfth ACM
  International Conference on Web Search and Data Mining}}.
  \bibinfo{pages}{519--527}.
\newblock


\bibitem[\protect\citeauthoryear{Ma, Zhang, Wang, and Liu}{Ma
  et~al\mbox{.}}{2018}]%
        {ma2018point}
\bibfield{author}{\bibinfo{person}{Chen Ma}, \bibinfo{person}{Yingxue Zhang},
  \bibinfo{person}{Qinglong Wang}, {and} \bibinfo{person}{Xue Liu}.}
  \bibinfo{year}{2018}\natexlab{}.
\newblock \showarticletitle{Point-of-Interest Recommendation: Exploiting
  Self-Attentive Autoencoders with Neighbor-Aware Influence}. In
  \bibinfo{booktitle}{\emph{Proceedings of the 27th ACM International
  Conference on Information and Knowledge Management}}. ACM,
  \bibinfo{pages}{697--706}.
\newblock


\bibitem[\protect\citeauthoryear{Massa and Avesani}{Massa and Avesani}{2007}]%
        {massa2007trust}
\bibfield{author}{\bibinfo{person}{Paolo Massa} {and} \bibinfo{person}{Paolo
  Avesani}.} \bibinfo{year}{2007}\natexlab{}.
\newblock \showarticletitle{Trust-aware recommender systems}. In
  \bibinfo{booktitle}{\emph{Proceedings of the 2007 ACM conference on
  Recommender systems}}. \bibinfo{pages}{17--24}.
\newblock


\bibitem[\protect\citeauthoryear{Ning and Karypis}{Ning and Karypis}{2011}]%
        {ning2011slim}
\bibfield{author}{\bibinfo{person}{Xia Ning} {and} \bibinfo{person}{George
  Karypis}.} \bibinfo{year}{2011}\natexlab{}.
\newblock \showarticletitle{Slim: Sparse linear methods for top-n recommender
  systems}. In \bibinfo{booktitle}{\emph{2011 11th IEEE International
  Conference on Data Mining}}. \bibinfo{pages}{497--506}.
\newblock


\bibitem[\protect\citeauthoryear{Quadrana, Karatzoglou, Hidasi, and
  Cremonesi}{Quadrana et~al\mbox{.}}{2017}]%
        {quadrana2017personalizing}
\bibfield{author}{\bibinfo{person}{Massimo Quadrana},
  \bibinfo{person}{Alexandros Karatzoglou}, \bibinfo{person}{Bal{\'a}zs
  Hidasi}, {and} \bibinfo{person}{Paolo Cremonesi}.}
  \bibinfo{year}{2017}\natexlab{}.
\newblock \showarticletitle{Personalizing session-based recommendations with
  hierarchical recurrent neural networks}. In
  \bibinfo{booktitle}{\emph{Proceedings of the Eleventh ACM Conference on
  Recommender Systems}}. \bibinfo{pages}{130--137}.
\newblock


\bibitem[\protect\citeauthoryear{Ram and Gray}{Ram and Gray}{2012}]%
        {ram2012maximum}
\bibfield{author}{\bibinfo{person}{Parikshit Ram} {and}
  \bibinfo{person}{Alexander~G Gray}.} \bibinfo{year}{2012}\natexlab{}.
\newblock \showarticletitle{Maximum inner-product search using cone trees}. In
  \bibinfo{booktitle}{\emph{Proceedings of the 18th ACM SIGKDD international
  conference on Knowledge discovery and data mining}}.
  \bibinfo{pages}{931--939}.
\newblock


\bibitem[\protect\citeauthoryear{Rendle, Freudenthaler, Gantner, and
  Schmidt-Thieme}{Rendle et~al\mbox{.}}{2009}]%
        {rendle2009bpr}
\bibfield{author}{\bibinfo{person}{Steffen Rendle}, \bibinfo{person}{Christoph
  Freudenthaler}, \bibinfo{person}{Zeno Gantner}, {and} \bibinfo{person}{Lars
  Schmidt-Thieme}.} \bibinfo{year}{2009}\natexlab{}.
\newblock \showarticletitle{BPR: Bayesian personalized ranking from implicit
  feedback}. In \bibinfo{booktitle}{\emph{Proceedings of the twenty-fifth
  conference on uncertainty in artificial intelligence}}.
  \bibinfo{pages}{452--461}.
\newblock


\bibitem[\protect\citeauthoryear{Rendle, Freudenthaler, and
  Schmidt-Thieme}{Rendle et~al\mbox{.}}{2010}]%
        {rendle2010factorizing}
\bibfield{author}{\bibinfo{person}{Steffen Rendle}, \bibinfo{person}{Christoph
  Freudenthaler}, {and} \bibinfo{person}{Lars Schmidt-Thieme}.}
  \bibinfo{year}{2010}\natexlab{}.
\newblock \showarticletitle{Factorizing personalized markov chains for
  next-basket recommendation}. In \bibinfo{booktitle}{\emph{Proceedings of the
  19th international conference on World wide web}}. \bibinfo{pages}{811--820}.
\newblock


\bibitem[\protect\citeauthoryear{Resnick, Iacovou, Suchak, Bergstrom, and
  Riedl}{Resnick et~al\mbox{.}}{1994}]%
        {resnick1994grouplens}
\bibfield{author}{\bibinfo{person}{Paul Resnick}, \bibinfo{person}{Neophytos
  Iacovou}, \bibinfo{person}{Mitesh Suchak}, \bibinfo{person}{Peter Bergstrom},
  {and} \bibinfo{person}{John Riedl}.} \bibinfo{year}{1994}\natexlab{}.
\newblock \showarticletitle{GroupLens: an open architecture for collaborative
  filtering of netnews}. In \bibinfo{booktitle}{\emph{Proceedings of the 1994
  ACM conference on Computer supported cooperative work}}.
  \bibinfo{pages}{175--186}.
\newblock


\bibitem[\protect\citeauthoryear{Sarwar, Karypis, Konstan, and Riedl}{Sarwar
  et~al\mbox{.}}{2001}]%
        {sarwar2001item}
\bibfield{author}{\bibinfo{person}{Badrul Sarwar}, \bibinfo{person}{George
  Karypis}, \bibinfo{person}{Joseph Konstan}, {and} \bibinfo{person}{John
  Riedl}.} \bibinfo{year}{2001}\natexlab{}.
\newblock \showarticletitle{Item-based collaborative filtering recommendation
  algorithms}. In \bibinfo{booktitle}{\emph{Proceedings of the 10th
  international conference on World Wide Web}}. \bibinfo{pages}{285--295}.
\newblock


\bibitem[\protect\citeauthoryear{Sedhain, Menon, Sanner, and Xie}{Sedhain
  et~al\mbox{.}}{2015}]%
        {sedhain2015autorec}
\bibfield{author}{\bibinfo{person}{Suvash Sedhain},
  \bibinfo{person}{Aditya~Krishna Menon}, \bibinfo{person}{Scott Sanner}, {and}
  \bibinfo{person}{Lexing Xie}.} \bibinfo{year}{2015}\natexlab{}.
\newblock \showarticletitle{Autorec: Autoencoders meet collaborative
  filtering}. In \bibinfo{booktitle}{\emph{Proceedings of the 24th
  International Conference on World Wide Web}}. \bibinfo{pages}{111--112}.
\newblock


\bibitem[\protect\citeauthoryear{Seo, Lin, Cohen, Shen, and Han}{Seo
  et~al\mbox{.}}{2016}]%
        {seo2016hierarchical}
\bibfield{author}{\bibinfo{person}{Paul~Hongsuck Seo}, \bibinfo{person}{Zhe
  Lin}, \bibinfo{person}{Scott Cohen}, \bibinfo{person}{Xiaohui Shen}, {and}
  \bibinfo{person}{Bohyung Han}.} \bibinfo{year}{2016}\natexlab{}.
\newblock \showarticletitle{Hierarchical attention networks}.
\newblock \bibinfo{journal}{\emph{arXiv preprint arXiv:1606.02393}}
  (\bibinfo{year}{2016}).
\newblock


\bibitem[\protect\citeauthoryear{Seo, Huang, Yang, and Liu}{Seo
  et~al\mbox{.}}{2017}]%
        {seo2017interpretable}
\bibfield{author}{\bibinfo{person}{Sungyong Seo}, \bibinfo{person}{Jing Huang},
  \bibinfo{person}{Hao Yang}, {and} \bibinfo{person}{Yan Liu}.}
  \bibinfo{year}{2017}\natexlab{}.
\newblock \showarticletitle{Interpretable convolutional neural networks with
  dual local and global attention for review rating prediction}. In
  \bibinfo{booktitle}{\emph{Proceedings of the Eleventh ACM Conference on
  Recommender Systems}}. \bibinfo{pages}{297--305}.
\newblock


\bibitem[\protect\citeauthoryear{Shrivastava and Li}{Shrivastava and
  Li}{2014}]%
        {shrivastava2014asymmetric}
\bibfield{author}{\bibinfo{person}{Anshumali Shrivastava} {and}
  \bibinfo{person}{Ping Li}.} \bibinfo{year}{2014}\natexlab{}.
\newblock \showarticletitle{Asymmetric LSH (ALSH) for sublinear time maximum
  inner product search (MIPS)}. In \bibinfo{booktitle}{\emph{Advances in Neural
  Information Processing Systems}}. \bibinfo{pages}{2321--2329}.
\newblock


\bibitem[\protect\citeauthoryear{Srivastava and Salakhutdinov}{Srivastava and
  Salakhutdinov}{2012}]%
        {srivastava2012multimodal}
\bibfield{author}{\bibinfo{person}{Nitish Srivastava} {and}
  \bibinfo{person}{Ruslan~R Salakhutdinov}.} \bibinfo{year}{2012}\natexlab{}.
\newblock \showarticletitle{Multimodal learning with deep boltzmann machines}.
  In \bibinfo{booktitle}{\emph{Advances in Neural Information Processing
  Systems}}. \bibinfo{pages}{2222--2230}.
\newblock


\bibitem[\protect\citeauthoryear{Sukhbaatar, Weston, Fergus,
  et~al\mbox{.}}{Sukhbaatar et~al\mbox{.}}{2015}]%
        {sukhbaatar2015end}
\bibfield{author}{\bibinfo{person}{Sainbayar Sukhbaatar},
  \bibinfo{person}{Jason Weston}, \bibinfo{person}{Rob Fergus},
  {et~al\mbox{.}}} \bibinfo{year}{2015}\natexlab{}.
\newblock \showarticletitle{End-to-end memory networks}. In
  \bibinfo{booktitle}{\emph{Advances in Neural Information Processing
  Systems}}. \bibinfo{pages}{2440--2448}.
\newblock


\bibitem[\protect\citeauthoryear{Tang and Wang}{Tang and Wang}{2018}]%
        {tang2018personalized}
\bibfield{author}{\bibinfo{person}{Jiaxi Tang} {and} \bibinfo{person}{Ke
  Wang}.} \bibinfo{year}{2018}\natexlab{}.
\newblock \showarticletitle{Personalized top-n sequential recommendation via
  convolutional sequence embedding}. In \bibinfo{booktitle}{\emph{Proceedings
  of the Eleventh ACM International Conference on Web Search and Data Mining}}.
  \bibinfo{pages}{565--573}.
\newblock


\bibitem[\protect\citeauthoryear{Tay, Anh~Tuan, and Hui}{Tay
  et~al\mbox{.}}{2018a}]%
        {tay2018latent}
\bibfield{author}{\bibinfo{person}{Yi Tay}, \bibinfo{person}{Luu Anh~Tuan},
  {and} \bibinfo{person}{Siu~Cheung Hui}.} \bibinfo{year}{2018}\natexlab{a}.
\newblock \showarticletitle{Latent relational metric learning via memory-based
  attention for collaborative ranking}. In
  \bibinfo{booktitle}{\emph{Proceedings of the 2018 World Wide Web Conference
  on World Wide Web}}. \bibinfo{pages}{729--739}.
\newblock


\bibitem[\protect\citeauthoryear{Tay, Tuan, and Hui}{Tay
  et~al\mbox{.}}{2018b}]%
        {tay2018multi}
\bibfield{author}{\bibinfo{person}{Yi Tay}, \bibinfo{person}{Luu~Anh Tuan},
  {and} \bibinfo{person}{Siu~Cheung Hui}.} \bibinfo{year}{2018}\natexlab{b}.
\newblock \showarticletitle{Multi-Pointer Co-Attention Networks for
  Recommendation}.
\newblock \bibinfo{journal}{\emph{arXiv preprint arXiv:1801.09251}}
  (\bibinfo{year}{2018}).
\newblock


\bibitem[\protect\citeauthoryear{Tran, Lee, Liao, and Lee}{Tran
  et~al\mbox{.}}{2018}]%
        {tran2018regularizing}
\bibfield{author}{\bibinfo{person}{Thanh Tran}, \bibinfo{person}{Kyumin Lee},
  \bibinfo{person}{Yiming Liao}, {and} \bibinfo{person}{Dongwon Lee}.}
  \bibinfo{year}{2018}\natexlab{}.
\newblock \showarticletitle{Regularizing Matrix Factorization with User and
  Item Embeddings for Recommendation}. In \bibinfo{booktitle}{\emph{Proceedings
  of the 27th ACM International Conference on Information and Knowledge
  Management}}. \bibinfo{pages}{687--696}.
\newblock


\bibitem[\protect\citeauthoryear{Van~den Oord, Dieleman, and Schrauwen}{Van~den
  Oord et~al\mbox{.}}{2013}]%
        {van2013deep}
\bibfield{author}{\bibinfo{person}{Aaron Van~den Oord}, \bibinfo{person}{Sander
  Dieleman}, {and} \bibinfo{person}{Benjamin Schrauwen}.}
  \bibinfo{year}{2013}\natexlab{}.
\newblock \showarticletitle{Deep content-based music recommendation}. In
  \bibinfo{booktitle}{\emph{Advances in Neural Information Processing
  Systems}}. \bibinfo{pages}{2643--2651}.
\newblock


\bibitem[\protect\citeauthoryear{Vaswani, Shazeer, Parmar, Uszkoreit, Jones,
  Gomez, Kaiser, and Polosukhin}{Vaswani et~al\mbox{.}}{2017}]%
        {vaswani2017attention}
\bibfield{author}{\bibinfo{person}{Ashish Vaswani}, \bibinfo{person}{Noam
  Shazeer}, \bibinfo{person}{Niki Parmar}, \bibinfo{person}{Jakob Uszkoreit},
  \bibinfo{person}{Llion Jones}, \bibinfo{person}{Aidan~N Gomez},
  \bibinfo{person}{{\L}ukasz Kaiser}, {and} \bibinfo{person}{Illia
  Polosukhin}.} \bibinfo{year}{2017}\natexlab{}.
\newblock \showarticletitle{Attention is all you need}. In
  \bibinfo{booktitle}{\emph{Advances in Neural Information Processing
  Systems}}. \bibinfo{pages}{6000--6010}.
\newblock


\bibitem[\protect\citeauthoryear{Wang, Guo, Lan, Xu, Wan, and Cheng}{Wang
  et~al\mbox{.}}{2015}]%
        {wang2015learning}
\bibfield{author}{\bibinfo{person}{Pengfei Wang}, \bibinfo{person}{Jiafeng
  Guo}, \bibinfo{person}{Yanyan Lan}, \bibinfo{person}{Jun Xu},
  \bibinfo{person}{Shengxian Wan}, {and} \bibinfo{person}{Xueqi Cheng}.}
  \bibinfo{year}{2015}\natexlab{}.
\newblock \showarticletitle{Learning hierarchical representation model for
  nextbasket recommendation}. In \bibinfo{booktitle}{\emph{ACM SIGIR Conference
  on Research and Development in Information Retrieval}}.
  \bibinfo{pages}{403--412}.
\newblock


\bibitem[\protect\citeauthoryear{Wu, Ahmed, Beutel, Smola, and Jing}{Wu
  et~al\mbox{.}}{2017}]%
        {wu2017recurrent}
\bibfield{author}{\bibinfo{person}{Chao-Yuan Wu}, \bibinfo{person}{Amr Ahmed},
  \bibinfo{person}{Alex Beutel}, \bibinfo{person}{Alexander~J Smola}, {and}
  \bibinfo{person}{How Jing}.} \bibinfo{year}{2017}\natexlab{}.
\newblock \showarticletitle{Recurrent recommender networks}. In
  \bibinfo{booktitle}{\emph{Proceedings of the tenth ACM international
  conference on web search and data mining}}. \bibinfo{pages}{495--503}.
\newblock


\bibitem[\protect\citeauthoryear{Wu, DuBois, Zheng, and Ester}{Wu
  et~al\mbox{.}}{2016}]%
        {wu2016collaborative}
\bibfield{author}{\bibinfo{person}{Yao Wu}, \bibinfo{person}{Christopher
  DuBois}, \bibinfo{person}{Alice~X Zheng}, {and} \bibinfo{person}{Martin
  Ester}.} \bibinfo{year}{2016}\natexlab{}.
\newblock \showarticletitle{Collaborative denoising auto-encoders for top-n
  recommender systems}. In \bibinfo{booktitle}{\emph{Proceedings of the Ninth
  ACM International Conference on Web Search and Data Mining}}.
  \bibinfo{pages}{153--162}.
\newblock


\bibitem[\protect\citeauthoryear{Xiong, Merity, and Socher}{Xiong
  et~al\mbox{.}}{2016}]%
        {xiong2016dynamic}
\bibfield{author}{\bibinfo{person}{Caiming Xiong}, \bibinfo{person}{Stephen
  Merity}, {and} \bibinfo{person}{Richard Socher}.}
  \bibinfo{year}{2016}\natexlab{}.
\newblock \showarticletitle{Dynamic memory networks for visual and textual
  question answering}. In \bibinfo{booktitle}{\emph{International Conference on
  Machine Learning}}. \bibinfo{pages}{2397--2406}.
\newblock


\bibitem[\protect\citeauthoryear{Xu, Ba, Kiros, Cho, Courville, Salakhudinov,
  Zemel, and Bengio}{Xu et~al\mbox{.}}{2015}]%
        {xu2015show}
\bibfield{author}{\bibinfo{person}{Kelvin Xu}, \bibinfo{person}{Jimmy Ba},
  \bibinfo{person}{Ryan Kiros}, \bibinfo{person}{Kyunghyun Cho},
  \bibinfo{person}{Aaron Courville}, \bibinfo{person}{Ruslan Salakhudinov},
  \bibinfo{person}{Rich Zemel}, {and} \bibinfo{person}{Yoshua Bengio}.}
  \bibinfo{year}{2015}\natexlab{}.
\newblock \showarticletitle{Show, attend and tell: Neural image caption
  generation with visual attention}. In \bibinfo{booktitle}{\emph{International
  Conference on Machine Learning}}. \bibinfo{pages}{2048--2057}.
\newblock


\bibitem[\protect\citeauthoryear{Xue, Dai, Zhang, Huang, and Chen}{Xue
  et~al\mbox{.}}{2017}]%
        {xue2017deep}
\bibfield{author}{\bibinfo{person}{Hong-Jian Xue}, \bibinfo{person}{Xinyu Dai},
  \bibinfo{person}{Jianbing Zhang}, \bibinfo{person}{Shujian Huang}, {and}
  \bibinfo{person}{Jiajun Chen}.} \bibinfo{year}{2017}\natexlab{}.
\newblock \showarticletitle{Deep Matrix Factorization Models for Recommender
  Systems}. In \bibinfo{booktitle}{\emph{Proceeding of the 26th International
  Joint Conference on Artificial Intelligence}}. \bibinfo{pages}{3203--3209}.
\newblock


\bibitem[\protect\citeauthoryear{Zhang, Yang, Luan, Yang, and Chua}{Zhang
  et~al\mbox{.}}{2014}]%
        {zhang2014start}
\bibfield{author}{\bibinfo{person}{Hanwang Zhang}, \bibinfo{person}{Yang Yang},
  \bibinfo{person}{Huanbo Luan}, \bibinfo{person}{Shuicheng Yang}, {and}
  \bibinfo{person}{Tat-Seng Chua}.} \bibinfo{year}{2014}\natexlab{}.
\newblock \showarticletitle{Start from scratch: Towards automatically
  identifying, modeling, and naming visual attributes}. In
  \bibinfo{booktitle}{\emph{Proceedings of the 22nd ACM International
  Conference on Multimedia}}. \bibinfo{pages}{187--196}.
\newblock


\bibitem[\protect\citeauthoryear{Zhou, Shan, Banerjee, and Sapiro}{Zhou
  et~al\mbox{.}}{2012}]%
        {zhou2012kernelized}
\bibfield{author}{\bibinfo{person}{Tinghui Zhou}, \bibinfo{person}{Hanhuai
  Shan}, \bibinfo{person}{Arindam Banerjee}, {and} \bibinfo{person}{Guillermo
  Sapiro}.} \bibinfo{year}{2012}\natexlab{}.
\newblock \showarticletitle{Kernelized probabilistic matrix factorization:
  Exploiting graphs and side information}. In
  \bibinfo{booktitle}{\emph{Proceedings of the 2012 SIAM International
  Conference on Data Mining}}. \bibinfo{pages}{403--414}.
\newblock


\end{thebibliography}
	
\end{document}